\documentclass[twocolumn, astrosymb, times,tighten, tikz,modern]{aastex631}

\usepackage{amsmath}
\usepackage{CJK}
\usepackage{graphicx}
\usepackage{hyperref}
\usepackage{enumerate}
\usepackage{subfigure}
\usepackage{makecell}
\usepackage{soul}
\usepackage{xcolor}

\begin{document}
\begin{CJK*}{UTF8}{gbsn}
\shortauthors{B.\ Peng, D.\ Valencia }

\title{Puffy Venuses: the Mass - Radius Impact of Carbon-Rich Atmospheres on Lava Worlds}

\author{Bo Peng (彭博)\email{bpeng@astro.utoronto.ca}}
\email{bpeng@astro.utoronto.ca}
\affiliation{Department of Astronomy \& Astrophysics, University of Toronto, 50 St. George St., Toronto, ON M5S 3H4, Canada}
\author{Diana Valencia\email{diana.valencia@utoronto.ca} }
\affiliation{Department of Astronomy \& Astrophysics, University of Toronto, 50 St. George St., Toronto, ON M5S 3H4, Canada}
\affiliation{Department of Physical and Earth Sciences, University of Toronto Scarborough, 1065 Military Trail, Toronto, ON, M1C 1A4, Canada}
\vspace{3cm}

\begin{abstract}

The recent advancements in exoplanet observations enable the potential detection of exo-Venuses, rocky planets with carbon-rich atmospheres. How extended these atmospheres can be, given high carbon abundances, has not been studied. To answer this, we present a model for a theoretical class of exoplanets - puffy Venuses - characterized by thick, carbon-dominated atmospheres in equilibrium with global magma oceans. Our model accounts for carbon and hydrogen partition between the atmosphere and the magma ocean, as well as the C-H-O equilibrium chemistry throughout a semi-grey, radiative-convective atmosphere. We find that radius inflation by puffy Venus atmospheres is significant on small and irradiated planets: carbon content of 1200 ppm (or that of ordinary chondrites) can generate an atmosphere of $\sim$ 0.16 - 0.3 $R_{\oplus}$ for an Earth-mass planet with equilibrium temperatures of 1500 to 2000 K. We identify TOI-561 b as an especially promising puffy Venus candidate, whose under-density could be attributed to a thick C-rich atmosphere. We also advocate for a puffy Venus interpretation of 55 Cancri e, where recent JWST observation indicates the presence of a CO/CO$_2$ atmosphere. Puffy Venuses may thus constitute a testable alternative interpretation for the interior structure of underdense low-mass exoplanets. 
\end{abstract}

\section{Introduction}\label{Sec:Intro} 

Observing outgassed secondary atmospheres offers an unparalleled window to understanding rocky exoplanet interiors and habitability. With the arrival of JWST, detecting such atmospheres on rocky exoplanets becomes feasible \citep{Ostberg19}.
Recent JWST phase-curve observation of the ultrashort period, underdense super-Earth 55 Cancri e indicate the exoplanet hosts a CO/CO$_2$ atmosphere \citep{Hu_2024_55CE_paper}. The emergence of a population of such underdense, highly irradiated rocky planets \citep{Piette_2023_Mixed_Atmosphere} motivates the question: can thick CO/CO$_2$ dominated atmospheres explain these planets' low density? What is the maximum extent of these Venus-like atmospheres, and can they create mass - radius (M-R) relationships distinct from that of Earth-like planets? 

Whether such thick carbon-rich atmospheres (i.e., ones dominated by C-bearing molecules) can exist hinges on the possible range of carbon abundances in rocky worlds. This is because their atmospheric height is primarily limited by their volatile budget, which is largely unconstrained. However, we can gain insight by considering two related questions - (a) if carbon-rich material is present in other exoplanetary systems, and (b) if the formation and atmospheric evolution pathways of these exoplanets can retain the carbon in their mantle and surface reservoirs. In the Solar System, chondritic samples suggest a bifurcation of carbon content in planetary building blocks - wet, carbonaceous chondrites (CC) beyond the snow line and dry, carbon-poor chondrites (NC) within it \citep{Bermingham20}. The most carbon-rich CI chondrites contain $\sim$3.5 wt$\%$ C \citep{Schaefer17}, which is $\sim$2 orders of magnitude higher than the bulk silicate Earth abundance \citep{bergin15, Fischer20}. Polluted white dwarf C/O data suggest that at least some exo-planetesimals have a similar bimodal distribution in carbon abundances \citep{Wilson_16_CO_planetesimal}. However, more exotic interior scenarios cannot be ruled out \citep{Putirka_21_polluted_WD_exotic_mineral}. 

Despite some caveats and unknowns, it is not difficult to envision a carbon-rich formation pathway for rocky exoplanets via pebble accretion, which efficiently grows planets from primitive material \citep{Ormel_10_first_pebble, Johansen_17_pebble_accretion_review, Johansen21, Izidoro_21_Super_Earth_formation_pebble}, thus avoiding some volatile loss pathways associated with hierarchical accretion \citep{Gu_2024_Earth_Volatiles}. A planet may receive volatile-rich pebbles drifting from beyond the snow line \citep{Sato_16_water_delivery_pebble, Ida_19_water_delivery_pebble}. Alternatively, it may form beyond the snow line rapidly, before the dispersal of the protoplanetary disc, and subsequently migrate inwards to become hotter - and thus puffier, making them easier to detect via transits \citep{Lambrechts19, Izidoro_21_Super_Earth_formation_pebble}. Potential caveats - mechanisms of carbon loss - include the atmospheric recycling with the surrounding disc \citep{Johansen21}, the dissolution and burial of carbon into the core \citep{Fischer20, Johansen_23_core_burial_atm_loss}, as well as the destruction of carbonaceous grains in the disc through photolysis \citep{LeeDisk, GTDust, Binkert_23_C_in_disc}. However, previous works on these mechanisms are primarily in the context of volatile delivery during Earth formation, which is under lively debate (see, e.g., \cite{Johansen21}). Therefore, their impact on the exoplanet population remains unclear. 

The above discussions suggest that carbon-rich exoplanets may exist. Yet, current exoplanet interior models have overlooked the potential significance of a thick carbon-dominated atmosphere in favor of hydrogen or water-rich layers (e.g., \citealt{Lacedelli_22_TOI561}), iron-depleted rocky interiors (e.g. \citealt{Plotnykov_Valencia_20_cmf}), or more exotic scenarios such as carbide worlds (e.g. \citealt{Madhu_2012_SiC_55CE}). In this study, we seek to understand the maximum impact of a carbon-dominated atmosphere on the radius of rocky exoplanets, a scenario we term puffy Venuses. Since we are interested in carbon-rich atmospheres as an alternative to H/He and steam atmospheres, we specifically probe atmospheres with C$\sim$O$>$H. We do this by constructing coupled atmosphere-magma ocean (MO) models since the thickest atmospheres would result in strong insulation and, thus, likely yield hot interiors where the mantle is substantially molten. We incorporate C-H-O thermochemistry, building on recent advances in modeling lava worlds in and beyond the Solar System \citep{Sossi_20_redox_MO_atmosphere, Bower22_COH_chem_MO, Gaillard_22_Redox_MO_degassing}. We demonstrate strong radius inflation among irradiated, low-mass puffy Venuses. For instance, a 0.16 - 0.3 $R_{\oplus}$ carbon-dominated atmosphere can exist on an Earth-mass, highly-irradiated ($T_{eq} \in [1500, 2000]$K) planet with modest (ordinary chondrite, 1200 ppm \citep{Schaefer17}) carbon abundances. We identify nine puffy Venus candidates and highlight our cases for TOI-561 b and 55 Cnc e.

\section{Magma Ocean - Atmosphere model}\label{Sec:Method}

We model planets with a carbon-rich atmosphere in thermochemical equilibrium with a magma ocean and obtain the height of the atmosphere $z_{atm}$. We distribute a C and H total volatile budget between a homogeneous MO and the atmosphere modeled via a 1-D semi-grey radiative-convective treatment (Section \ref{apdx:dissolution}), which allows for the development of deep radiative zones that  \citet{Selsis_2023} demonstrated to be important for thick secondary atmospheres. We find $z_{atm}$ as the height where the atmosphere thermal radiation reaches a transit optical depth of 0.56 \citep{deWit_Seager_2013_transmit_tau, Heng_Kitzmann_2017_Transmit_tau}.

We consider an atmosphere composed of 6 species - CO, CO$_2$, CH$_4$, H$_2$, H$_2$O, and O$_2$ (Section \ref{apdx:CE}). We recognize that besides the volatile C and H species, these exoplanets can also contain C and H species not available to the atmosphere, such as those dissolved into the metallic core. But their abundance is likely limited (Section \ref{apdx:caveats_other_C}). We consider the atmosphere to be either (I) in chemical equilibrium (CE) at each layer and in CE with the MO at the interface or (II) well-mixed with a composition set by the CE with the magma ocean at the interface. Case I represents the scenario when chemical kinetics is faster than vertical mixing, while case II represents the reverse. We also consider (III) pure CO$_2$-H$_2$O atmospheres as an end member case. In all scenarios, the MO and the atmosphere are coupled at their interface via solubility laws (see Section \ref{apdx:dissolution}). We find the mass of the MO by calculating the adiabatic mantle thermal profile assuming an Earth-like lower-mantle petrology.

\subsection{Atmospheric chemical equilibrium \& equations of state}\label{apdx:CE}

We use three gas-phase chemical reactions to couple the six volatile species in our atmosphere model - $\mathrm{H_2}$, $\mathrm{O_2}$, $\mathrm{H_2O}$, $\mathrm{CO_2}$, $\mathrm{CO}$ and $\mathrm{CH_4}$:

\begin{equation}\label{eqn:A1}
    \mathrm{CO + 0.5\,O_2} \Leftrightarrow \mathrm{CO_2}
\end{equation}
\begin{equation}
    \mathrm{H_2 + 0.5\,O_2} \Leftrightarrow \mathrm{H_2O}
\end{equation}
\begin{equation}
    \mathrm{CO_2 + 2\,H_2}  \Leftrightarrow \mathrm{CH_4 + O_2}
\end{equation}

As detailed below, the equilibrium molar fractions of these gases, $y_i$'s, are solved given total pressure (P), temperature (T), and atomic ratios C/O and C/H. In the fiducial Scenario I, we use the local P and T throughout the radiative-convective atmosphere, which is obtained from the numerical integration of the structure equations (see Section \ref{Subsec:atm_prof}). We ensure convergence given that the molar fraction of gases feed back into the structure equations via density and mean specific heat capacity. In Scenario II, instead, we only solve for CE $y_i$'s at the magma ocean surface P-T. The chemical equilibrium between the atmosphere and the magma ocean is handled by the Henrian solubility laws using $y_i$'s and the surface pressure, see Section \ref{apdx:dissolution}.

Each reaction corresponds to an equilibrium constant $K_{eq}$:

\begin{equation}\label{eqn:K}
    K_{eq, 1}(T) = \frac{f_{\mathrm{CO}}f_{\mathrm{O_2}}^{\,\,0.5}}{f_{\mathrm{CO_2}}},
\end{equation}
\begin{equation}
    K_{eq, 2}(T) = \frac{f_{\mathrm{H_2}}f_{\mathrm{O_2}}^{\,\,0.5}}{f_{\mathrm{H_2O}}},
\end{equation}
\begin{equation}
    K_{eq, 3}(T) = \frac{f_{\mathrm{CO_2}}f_{\mathrm{H_2}}^{\,\,2}}{f_{\mathrm{CH_4}}f_{\mathrm{O_2}}},
\end{equation}
where $f_{i}$ denotes the fugacity of gas species $i$. Fugacity describes the availability of a compound for chemical reactions. We calculate $K_{eq}(T)$ utilizing the thermochemical data from the NIST database \citep{NIST}. As numerical fits to the Shomate equation, the NIST data are accurate up to 6000 K. We linearly extrapolate the NIST data beyond 6000 K, but we note that our fiducial models predict deep atmosphere temperatures $\lesssim 7000$ K (see fig. \ref{fig:atm_prof}d). We note that higher temperature behaviors of $K_{eq}$'s remain qualitatively sensible. 

In general, a real gas $i$'s fugacity is a function of both the local pressure and temperature, and the composition of the gas mixture. In other words, $f_i = f(T, P, y_i, y_{j\neq i})$, where $y_i = N_i/N$ is the molar fraction.
In practice, comparable atmosphere speciation models often assume ideal gas, $f_i = y_i P$ \citep{Bower22_COH_chem_MO, Shorttle_2024_MO_K2_18b}, or non-ideal gases in ideal mixture $f_i = f(T, P, y_i)$ (e.g., \citealt{Sun_Lee_22_MO_Redox}) to simplify the calculation, with the notable exception of \cite{Tian_Heng_2024_Hybrid_atm}, who incorporated non-ideal mixing for a subset of their species. We choose to adopt the non-ideal gas in ideal mixture approach to balance realism and tractability. We discuss this choice further in Section \ref{apdx:caveats_MO_surf}. 

Under our assumption, fugacity and local P-T are related by 

\begin{equation}
    f_i = \phi_{i, pure}y_{i} P,
\end{equation}
where $\phi_{i, pure}$ is the fugacity coefficient calculated from the equation of state (EoS) of pure gas $i$. Here, P is the total pressure at a radial location in the atmosphere. We use the modified Lee and Kesler EoS developed by \cite{Duan_1992_EOS} for CO$\mathrm{_2}$ and H$_2$O, and use a generalized version of it for the other 4 species \citep{Duan_1996_EOS}. The \cite{Duan_1992_EOS} EoS is accurate to 1273 K and 8000 bars and approximately accurate for higher P-T, while the \cite{Duan_1996_EOS} EoS is accurate to 2000K and 25000 bars. The equation of state is expressed in terms of compressibility $Z_i$

\begin{equation}
    Z_i(P, T) \equiv \frac{PV}{R_{gas}T}, 
\end{equation}

where $V$ is the molar volume and $R_{gas}$ is the gas constant. Then from first principles, $\phi_{i, pure}$ can be found via \citep{Duan_1992_EOS}

\begin{equation}
    \ln \phi_{i, pure}(P, T) = \int_0^P (Z(P, T) - 1)\frac{\mathrm{d}P'}{P'}.
\end{equation}

In practice, we use the EoS and the analytical $\phi(Z, P, T)$ relationship from \cite{Duan_1992_EOS} to calculate $Z_i$ and $\phi_i$ tables. We then interpolate from these tables during atmosphere integration. We assume constant atomic C/O and C/H at each layer of the atmosphere 
\begin{equation}
    \mathrm{C/O} = \frac{y_{\mathrm{CO_2}} + y_{\mathrm{CO}} +  y_{\mathrm{CH_4}}}{2y_{\mathrm{CO_2}} + y_{\mathrm{CO}} + 2y_{\mathrm{O_2}} + y_{\mathrm{H_2O}}},
\end{equation}
\begin{equation}\label{eqn:C/H}
    \mathrm{C/H} = \frac{y_{\mathrm{CO_2}} + y_{\mathrm{CO}} +  y_{\mathrm{CH_4}}}
    {4y_{\mathrm{CH4}} + 2y_{\mathrm{H_2O}} +  2y_{\mathrm{H_2}}}
\end{equation}

Finally, we close the system by enforcing $\sum y_i = 1$, and solve \ref{eqn:K} through \ref{eqn:C/H} numerically for the local gas mixture in equilibrium: $y_i(P, T, \mathrm{C/O}, \mathrm{C/H})$. 

The C/O and C/H ratios are set by the MO-atmosphere interaction. C/O is a proxy for the atmosphere's redox state. Because the mantle has vastly more redox power than the outgassed atmosphere, the former should control the latter's redox state. In the literature, the mantle redox state is usually parameterized by mineral buffers, most commonly the iron-wustite buffer (e.g., \citealt{Gaillard_22_Redox_MO_degassing, Bower22_COH_chem_MO, Tian_Heng_2024_Hybrid_atm}). However, these buffers are experimentally calibrated to Earth mantle P-T, below $\sim$ 3000K \citep{Ballhaus_1991_IW, Hirschmann_21_IW}. This is easily exceeded in our deep atmospheres, which reach $>$6000 K for the most irradiated planets. We, therefore, choose to specify the C/O for our atmospheres as an input variable. 

Although this approach effectively divorces the redox of the atmosphere from mantle geochemistry, it has merits beyond necessity. Firstly, how redox reactions such as IW, active in the deep mantle, influence the oxidation state at the surface of the magma ocean is complex \citep{HIRSCHMANN2012, Hirschmann_22_MO_Fe_Cr_redox, Gaillard_22_Redox_MO_degassing}. Further, the extent to which extrasolar magma oceans share similar composition and geochemistry as their terrestrial counterparts is uncertain. In fact, greater diversity in exoplanet composition has been inferred from polluted white dwarfs \citep{Putirka_21_polluted_WD_exotic_mineral}. Moreover, the C/O ratio of an exoplanet atmosphere is potentially observable in the JWST era and has started to be constrained for some hot Jupiters \citep{Line_21_CO_HJ_measurment, Changeat_22_25HJ_atm} and sub-Neptunes \citep{Benneke_2024_TOI_270d}.

To summarize, chemical speciation is calculated via solving \ref{eqn:K} through \ref{eqn:C/H}, using P, T, C/O and C/H as input. For Scenario I, integrating atmospheric profile provides local $P(r)$ and $T(r)$, while Scenario II only requires MO surface P and T. C/O is a model input, while C/H is iteratively found via the MO-atmosphere mass balance that we describe in Section \ref{apdx:dissolution}.

\subsection{1-D radiative-convective atmosphere profile}\label{Subsec:atm_prof}

We developed a simple 1-D radiative-convective model based on the semi-gray treatment of \cite{Guillot_2010}, hereafter G10. This scheme treats the planet's thermal emission and stellar irradiation separately, each characterized by a mean opacity. This scheme allows us to account for a deep radiative zone that a long-living lava-world atmosphere likely develops \citep{Selsis_2023}. 

We choose this treatment over a full-physics radiative-convective atmosphere model both for the simple model's conceptual clarity and for necessity. Since our model is concerned with constraining the atmospheric thickness for the population of C-rich exoplanets rather than predicting in detail the spectral signatures of specific exoplanets, our simple treatment suffices. Moreover, while recent efforts (e.g., \citealt{HITRAN_2022}) have expanded our understanding of the relevant molecules' radiative properties up to P $\leq$ 1000 bars and T $\lesssim$ 4000 K, the deep atmospheres we probe (T $\gtrsim$ 3000 K, P $ \gtrsim$ 3000 bars) remains elusive. Moreover, the cross-species (e.g., CO-CO$\mathrm{_2}$, CO$_2$-O$_2$) collision-induced absorption is not well-characterized to high temperatures (see review in \citealt{Chubb_review_atm_data}). 

We also note that G10's treatment has been further developed by, e.g., \cite{Heng_2012_G10, Heng_2014_G10, Parmentier_2014_update_G10} to account for the effects of scattering clouds and hazes and variable optical and infrared opacities. While these fruitful efforts increase the realism and nuance of this approach, we choose to use G10's version for its minimal number of parameters. We note that our approach is still an advancement compared to previous magma ocean-atmosphere models that assume 0-D or adiabatic atmospheres (e.g. \citealt{Gaillard_22_Redox_MO_degassing, Lichtenberg_21_Solubility, Kat21, Bower22_COH_chem_MO}, with the exceptions of \citealt{Selsis_2023, Shorttle_2024_MO_K2_18b}). 

Overall, our model integrates the atmospheric pressure-temperature profile using the differential form of G10's system of equations for isotropic irradiation and switching to a dry, non-ideal adiabat when appropriate. We also account for atmosphere self-gravity. For Scenario I, local gas composition $y_i$ is informed by thermochemical equilibrium. $y_i$ then feeds back to the adiabatic lapse rate and density profile. For Scenario II, the atmospheric profile takes constant $y_i$ throughout the atmosphere. See Section \ref{apdx:dissolution} for details.

\subsubsection{Radiative Temperature Gradient}\label{subsec:Rad_T_gradient}

We first derive the radiative temperature gradient following G10. Here we summarize our treatment. We start with the first 3 Eddington moments of the intensity $I_{\nu\mu}$, defined as - 
\begin{equation}
    (J_{\nu}, H_{\nu}, K_{\nu}) = \frac{1}{2}\int_{-1}^{1} I_{\nu\mu} (1, \mu, \mu^2)\mathrm{d}\mu
\end{equation}
where $J_{\nu}$, $4\pi H_{\nu}$, and $4\pi K_{\nu}/c$ corresponds to the energy of the beam, the radiation flux, and radiation pressure, respectively. $\mu = \cos \theta$ is the incident angle. Assuming the incoming visible, and outgoing thermal radiation fields are independent, their wavelength-integrated moments become   

\begin{equation}
    (J_v, H_v, K_v) \equiv \int_{visible} (J_{\nu}, H_{\nu}, K_{\nu})\mathrm{d}\nu
\end{equation} 
\begin{equation}
    (J_{th}, H_{th}, K_{th}) \equiv \int_{thermal} (J_{\nu}, H_{\nu}, K_{\nu})\mathrm{d}\nu
\end{equation}

In a planar atmosphere in thermodynamic equilibrium, these moments of radiation are related by

\begin{equation}\label{eqn:dH_v=J_v}
    \frac{\mathrm{d}H_v}{\mathrm{d}m} = \kappa_v J_v
\end{equation} 

\begin{equation}\label{eqn:dK_v=H_v}
     \frac{\mathrm{d}K_v}{\mathrm{d}m} = \kappa_v H_v
\end{equation}

\begin{equation}\label{eqn:dH_th=J_th}
    \frac{\mathrm{d}H_{th}}{\mathrm{d}m} = \kappa_{th} (J_{th} - B)
\end{equation} 

\begin{equation}\label{eqn:dK_th=H_th}
     \frac{\mathrm{d}K_{th}}{\mathrm{d}m} = \kappa_{th} H_{th},
\end{equation}

where $ B = \sigma/\pi T^4$ is the local black-body radiance, and $\sigma$ is the Stefan-Boltzmann constant. $m = -\int\rho \mathrm{d}r$ is the atmosphere column mass, integrating from the ``top" of the atmosphere downwards. The local mean gas density $\rho$  at the radial location $r$ is found using the non-ideal EoS
\begin{equation}\label{eqn:rho_NI}
    \rho = \frac{P}{R_{gas}T} \frac{\sum y_i \mu_i}{\sum y_i Z_i},
\end{equation}
where $\mu_i$ is the molecular mass of species $i$. Because $Z\geq1$ in our P-T range, our non-ideal atmosphere tends to be less dense and puffier than their ideal counterparts. 

$\kappa_{v}$ and $\kappa_{th}$ are mean opacities  

\begin{equation}
    \kappa_{v} = J_{v}^{-1}\int_{v}\kappa_{\nu}J_{\nu} \mathrm{d}\nu
\end{equation}

\begin{equation}
    \kappa_{th} = J_{th}^{-1}\int_{th}\kappa_{\nu}J_{\nu} \mathrm{d}\nu
\end{equation}

$\kappa_{v}$ and $\kappa_{th}$ are complex functions of atmospheric composition and stellar irradiation. Our model allows for any functional form of visible and thermal absorption coefficients $\kappa_{v, th}(P,T)$. But we, by default, use constant $\kappa$'s. We choose conservative values that fit the P-T observations of the Venusian atmosphere: $\kappa_{v} = 7.524\cdot 10^{-6} \mathrm{cm^2/g}$ and $\kappa_{th} = 3.3\cdot 10^{-3} \mathrm{cm^2/g}$. We also tested  $\kappa \propto P$, which is motivated by the pressure-broadening of collision-induced absorption lines \citep{Heng_2012_G10, Tolento_2019_analytical_atm}. The two cases produce similar results, so we choose the simpler constant case. We discuss the caveats of this treatment and show that our main result is robust against our opacity choices in Section \ref{apdx:opacity}.

We can then define optical depths for both radiation fields
\begin{equation}\label{eqn:dtau}
    \mathrm{d}\tau_{v, th} = \kappa_{v, th}\mathrm{d}m
\end{equation}

Local thermodynamic equilibrium dictates 
\begin{equation}\label{eqn:LTE_condition}
    0 = \kappa_{th} (J_{th} - B) + \kappa_{v}J_{v}
\end{equation}

This is equivalent to writing $H_{v} + H_{th} \equiv H_{int}$, where $H_{int} = \sigma T_{int}^4/(4\pi)$ is the outgoing heat flux at the bottom of the atmosphere, or the internal heat flux\footnote{We assume no surface effect as our atmosphere is thick}. Geophysically speaking, $H_{int}$ roots from the residual heat of planet formation and the radiogenic heating of the planet, encoding its thermal history. We choose $T_{int} = 0$ K as the fiducial value to model a planet at thermal equilibrium, yielding the least amount of inflation with no additional heat warming up the deep atmosphere. Choosing Earth or Jupiter-level internal heating only marginally inflates the atmosphere since the high stellar irradiation dominates a puffy Venus's energy budget.

To close the system of equations, we invoke the Eddington approximation following G10 
\begin{equation}\label{eqn:K_th = 3J_th}
    K_{th} = \frac{1}{3} J_{th}
\end{equation} 
\begin{equation}\label{eqn:K_v = 3J_v}
    K_{v} = \frac{1}{3} J_{v}
\end{equation}

They are strictly correct for an isotropic atmosphere or in the two-stream approximation and are generally accurate to the first order of magnitude \citep{Parmentier_2014_update_G10}. For the incoming radiation field, combining \ref{eqn:dH_v=J_v},\ref{eqn:dK_v=H_v}, and \ref{eqn:K_v = 3J_v} gives
\begin{equation}
    (J_{v}(\tau_v), H_{v}(\tau_{v})) = (J_{v0}, H_{v 0})\exp(-\sqrt{3}\tau_{v}),
\end{equation}
where $J_{v0} = - \sqrt{3} H_{v0}$ are outer boundary conditions at $\tau = 0$.  $H_{v0} = - \sigma T_{eq}^4/(\sqrt{3}\pi)$ is the incoming stellar radiation, which is a downward flux and, thus, negative.

We define the opacity ratio folloiwng G10 as $\gamma \equiv \kappa_{v}/\kappa_{th}$. Rewriting Equation \ref{eqn:LTE_condition} in terms of $J_{th}$ and taking its derivative, we have
\begin{equation}\label{eqn:dJ_th}
    \frac{\mathrm{d} J_{th}}{\mathrm{d}m} = \frac{\mathrm{d}B}{\mathrm{d}m} - \frac{\mathrm{d}}{\mathrm{d}m}(\gamma J_v)
\end{equation}
By combining the above with Equations \ref{eqn:dK_th=H_th} and \ref{eqn:K_th = 3J_th}, we obtain
\begin{equation}
    \kappa_{th} H_{th} = \frac{1}{3}\Big[
    \frac{\mathrm{d}B}{\mathrm{d}m} - \Big(\frac{\mathrm{d}\gamma}{\mathrm{d}m}J_v + \frac{\mathrm{d}J_v}{\mathrm{d}m} \gamma \Big)\Big].
\end{equation}
We express the P and T dependencies for $\gamma$ and T dependency of $B$ explicitly, and use Equations \ref{eqn:dK_v=H_v} and \ref{eqn:K_v = 3J_v} to rewrite the last term, obtaining 

\begin{equation}\label{eqn:one_step_before_rad_grad}
    \begin{split}
        0 = & \frac{1}{3}\frac{\mathrm{d}B}{\mathrm{d}T}\frac{\mathrm{d}T}{\mathrm{d}m} \\
        &- \frac{1}{3} \Big(\frac{\partial \gamma}{\partial T}\frac{\mathrm{d}T}{\mathrm{d}m} J_{v} + \frac{\partial \gamma}{\partial P}\frac{\mathrm{d}P}{\mathrm{d}m} J_{v}\Big) \\
        &- (\kappa_v\gamma H_v + \kappa_{th} H_{th}).
    \end{split}
\end{equation}

We enforce hydrostatic equilibrium 
\begin{equation}\label{eqn:hydrostat}
    \mathrm{d}P = g(r)\mathrm{d}m = - \rho(r) g(r) \mathrm{d}r,
\end{equation}

and rearrange Equation \ref{eqn:one_step_before_rad_grad}, using the definition of $B$ and switching variables $\mathrm{d}r = -1/\rho\cdot \mathrm{d}m$ to obtain the radiative temperature gradient 
\begin{equation}\label{eqn:dT_rad}
    \begin{split}
        \frac{\mathrm{d}T}{\mathrm{d}r}\Big|_{rad} = - &\rho 
    \Big[ \frac{\partial \gamma}{\partial P} g(r) J_{v}\\
    &+ 3(\kappa_v\gamma H_v + \kappa_{th} H_{th})\Big]\\
    & \times
    \Big(\frac{4\sigma}{\pi}T^3 - \frac{\partial \gamma}{\partial T}J_v \Big)^{-1}
    \end{split}
\end{equation}

Here $g(r) = GM(r)/r^2$ is the local gravitational acceleration accounting for atmosphere self-gravity, and $M(r)$ is the mass of the planet enclosed at radius $r$, which we numerically integrate via
\begin{equation}\label{eqn:dM}
    \mathrm{d}M = 4\pi r^2 \rho (r) \mathrm{d}r,
\end{equation}

\vspace{0.5cm}

\subsubsection{Non-ideal Adiabat}

When the radiative gradient exceeds the adiabat, the atmosphere profile follows a non-ideal adiabat - adiabat generalized to real-gas EoS. The adiabat is given by 
\begin{equation}\label{eqn:dT_adb}
    \frac{\mathrm{d}T}{\mathrm{d}P}\Big|_{ad} = \frac{T}{\overline{c_p}}   \frac{\mathrm{d}V}{\mathrm{d}T}\Big|_{P},
\end{equation}
where $\overline{c_p}$ is the mean specific heat. We calculate $\overline{c_p}(y_i, T)$ via

\begin{equation}
    \overline{c_p}(T) = \sum_{i} y_i c_{p, i}(T).
\end{equation}

Specific heat for individual species, $c_{p, i}(T)$, are calculated using a Shomate equation fit to tabulated data provided by NIST \citep{NIST}
\begin{equation*}
    c_{p, i}(T) = A + B\cdot t + C \cdot t^2 + D\cdot t^3 + E\cdot^{-2},
\end{equation*}
where $A$, $B$, $C$, $D$ and $E$ are compound-specific fitting constants and $t = T/1000$. 
We linearly extrapolate $c_{p, i}(T)$ beyond 6000 K. We calculate the volume gradient for an individual gas from the EoS
\begin{equation}
    \frac{\mathrm{d}V_i}{\mathrm{d}T}\Big|_{P} = Z_i\frac{R_{gas}}{P} + \frac{\mathrm{d}Z_i}{\mathrm{d}T}\Big|_{P}
    \frac{R_{gas} T}{P}
\end{equation}
where, like $Z_i(P, T)$ and $\phi_{i}(P, T)$, we pre-calculate $\frac{\mathrm{d}Z_i}{\mathrm{d}T}\Big|_{P}$ tables from the EoS and interpolate with them when integrating the atmosphere profile. The mean $\frac{\mathrm{d}V}{\mathrm{d}T}\Big|_{P}$ is found via -
\begin{equation}
    \frac{\mathrm{d}V}{\mathrm{d}T}\Big|_{P} = \sum y_i \frac{\mathrm{d}V_i}{\mathrm{d}T}\Big|_{P} + \sum V_i \frac{\mathrm{d}y_i}{\mathrm{d}T}\Big|_{P}.
\end{equation}
The second term accounts for the shift of the equilibrium gas mixture in response to temperature change. The non-ideality in our adiabat has a minor effect on the atmospheric height, inflating the atmosphere by a couple of percent.

\subsubsection{Boundary conditions \& transit height}
We integrate the atmospheric profiles numerically from the top $r_{top}$ to the MO surface at $R_s$, using the {\fontfamily{qcr}\selectfont LSODA} integrator in the {\fontfamily{qcr}\selectfont SciPy} package \citep{SciPy_20}. Specifically, we integrate \ref{eqn:hydrostat} for $P$, equations \ref{eqn:dT_rad} and \ref{eqn:dT_adb} for $T$, \ref{eqn:dM} for $M(r)$, and \ref{eqn:dtau} for $\tau_v$ and $\tau_{th}$. At the top of the atmosphere, we choose $P_{top} = 0.1$ Pa, $\tau_{v, top} = 10^{-9}$ and $\tau_{th, top} = \tau_{th, top}/\gamma$. We choose fiducial $\gamma = 2.28\cdot 10^{-3}$ (see Section \ref{apdx:opacity}). The top temperature is given by Equation 28 of G10, taking their incident angle $\mu_{*} = 1/\sqrt{3}$ and their irradiation temperature $T_{irr}^4 = 4T_{eq}^4$: 

\begin{equation}
    T_{top} = \Big[ \frac{1}{2}T_{int}^4 + \frac{2}{\sqrt{3}}T_{eq}^4
    \Big(1 + \frac{\sqrt{3}}{2}\gamma\Big) \Big]
\end{equation}

We numerically integrate the transit, or chord, optical depth (see G10) in the thermal wavelength $\tau_{th, ch}$, and define the height of the atmosphere $z_{atm}$ as the location where $\tau_{th, ch} = 0.56$, a number derived for transiting isothermal atmospheres from first principles \citep{deWit_Seager_2013_transmit_tau, Heng_Kitzmann_2017_Transmit_tau}. Since the high atmosphere in G10 is close to isothermal, this is an appropriate choice. Choosing the visible transit location $\tau_{v, ch} = 0.56$, or the commonly-used $\tau_{th, ch} = 2/3$ as the criteria do not strongly impact our results. 

From integrating the atmospheric profile, we also obtain the total mass of the atmosphere $M_{atm}$, as well as the total mass of C and H in the atmosphere $M_{C, H, atm}$ for the mass balance calculation (section \ref{apdx:dissolution}).

\subsection{Thermal structure of the mantle \& the magma ocean depth}\label{apdx:interior_model}

Our mantle model calculates the thermal structure and finds the thickness of the molten layer. Given $M_s$ as input, we find $R_s$, the radius of the planet's silicate-iron interior, by interpolating the mass-radius data from a detailed super-Earth interior model by \citet{Plotnykov_Valencia_20_cmf}, which assumes an Earth-like core-mass fraction of 0.325. However we note that assuming a power-law relation $R_s/R_{\oplus} = (M_s/M_{\oplus})^{0.27}$ only slightly impacts our results.

We integrate the adiabatic mantle temperature profile $T(r, \rho)$ using the Vinet EoS for the lower mantle, a mixture of bridgmanite and magnesiowustite \citep{Plotnykov_Valencia_20_cmf}. We discuss the inaccuracies of this simplified approach in Section \ref{apdx:caveats_mantle_thermal_model} and deem them acceptable. 

To find the extent of mantle melting, we combine two sets of solidus $T_{sol} (P)$ and liquidus $T_{liq} (P)$ curves. We use peridotite curves from \cite{MORSCHHAUSER_2011_UM_Melting} for the low-pressure end and chondritic curves from \cite{Montreux_16_deep_MO_melting} up to 140 GPa. Beyond that, we linearly interpolate the \cite{Montreux_16_deep_MO_melting} curves. We find the local melt fraction via linear interpolation 

\begin{equation}
    \phi_{melt} = \frac{T - T_{sol}}{ T_{liq} - T_{sol}}.
\end{equation}

We obtain the depth of the magma ocean at the location where $r(\phi_{melt} = 0.4)$ when the rock-melt mixture transition between rock-like to liquid-like viscosity \citep{Abe_93_rheological_transition}. We integrate the mass of the planet above this depth to find the MO mass $M_{MO}$.

\subsection{MO-atmosphere C \& H mass balance}\label{apdx:dissolution}

To account for the chemical exchange between the atmosphere and magma ocean we assume vigorous mixing to bring the atmosphere and the magma ocean to equilibrium at their interface, similar to previous magma ocean models (e.g \citealt{Bower22_COH_chem_MO, Lichtenberg_21_Solubility, Gaillard_22_Redox_MO_degassing}). This means that we set the bulk magma ocean C and H abundances by the chemical equilibrium and solubility relations at the MO surface. We describe the partitioning of the dissolved vs. outgassed gas species with the Henrian solubility relations from \citet{Lichtenberg_21_Solubility}. These relations link the partial pressures of individual species to their dissolved abundances in the MO:
\begin{equation}\label{eqn:solubility}
    X_i = a \cdot (y_{i, s} P_s) ^{b},
\end{equation}
where $X_i$ is the abundance of dissolved gas $i$ in the MO by weight, and $a$ and $b$ are constant fits to experimental data. $P_s$ is the total pressure, and $y_{i,s}$ is the molar fraction, both at the magma ocean surface. As we describe later, these are from the numerically integrated atmospheric profile.

We find the total dissolved elemental C and H mass, $M_{C,\, MO}$ and $M_{H,\, MO}$ via bookkeeping 

\begin{equation}
    M_{C,\, MO} = M_{MO} \sum_i k_{i, C}\frac{\mu_C}{\mu_i}X_i
\end{equation}
\begin{equation}\label{eqn:MO_volatiles}
    M_{H,\, MO} = M_{MO} \sum_i k_{i, H}\frac{\mu_H}{\mu_i}X_i,
\end{equation}

where $k_{i, C}$ and $k_{i, H}$ are the number of C and H atoms in compound $i$. $\mu_{C}$, $\mu_H$ and $\mu_i$ are molar weights of C, H and species $i$. The total C and H abundances ($x_{C, tot}$, $x_{H, tot}$) then found via 
\begin{equation}\label{eqn:x_tot}
    x_{C,  tot} = \frac{1}{M_{tot}}(M_{C, MO} + M_{C,  atm}).
\end{equation}
\begin{equation}\label{eqn:x_tot2}
    x_{H, tot} = \frac{1}{M_{tot}}(M_{H, MO} + M_{H, atm}).
\end{equation}

Overall, the coupled atmosphere magma ocean model takes $x_{C, tot}$, $x_{H, tot}$, $M_s$, $R_s$, and atmospheric C/O as inputs. Numerically, we start with guesses of the top-of-atmosphere (ToA) conditions $r_{top}$ and $M_{tot}$, and in Scenario I, the atmospheric C/H. We integrate from $r_{top}$ to the MO surface at $R_s$ to find the MO surface conditions $P_s$, $y_{i,s}$ and $T_s$, as well as $M(r = R_s)$, the mass enclosed by the MO surface from integration (Sections \ref{apdx:CE} and \ref{Subsec:atm_prof}). The first three of these then inform the MO volatile abundance $M_{C, H, MO}$ via equations \ref{eqn:solubility} to \ref{eqn:MO_volatiles}.  Together with the atmospheric C and H masses $M_{C, H, atm}$, we find bulk abundances $x_{C, tot}$ and $x_{H, tot}$. We then iterate
the ToA $r_{top}$ and $M_{tot}$, as well as atmospheric C/H until we find the atmosphere-MO pair that produces the desired $x_{C, tot}$ and $x_{H, tot}$, as well as the self-consistent requirement that the mass enclosed at the MO surface from atmosphere integration equals the mass of the rocky planet below 

\begin{equation}\label{eqn:M_s_requirement}
    M(r = R_s) = M_s.
\end{equation}

In the chemically quenched Scenario II, in addition to $M_{tot}$, $r_{top}$ and C/H, the atmosphere model also iterates for the MO surface conditions $P_s$ and $T_s$ to find the correct gas mixture $y_i$ that satisfies the aforementioned requirements. To close the system, we also enforce self-consistency on the atmosphere P-T profile - 
\begin{equation}
    (P(r = R_s), T(r = R_s))  = (P_s, T_s).
\end{equation}

\section{Results}\label{Sec:Results}

\begin{figure}
    \centering
    \includegraphics[width =\linewidth]{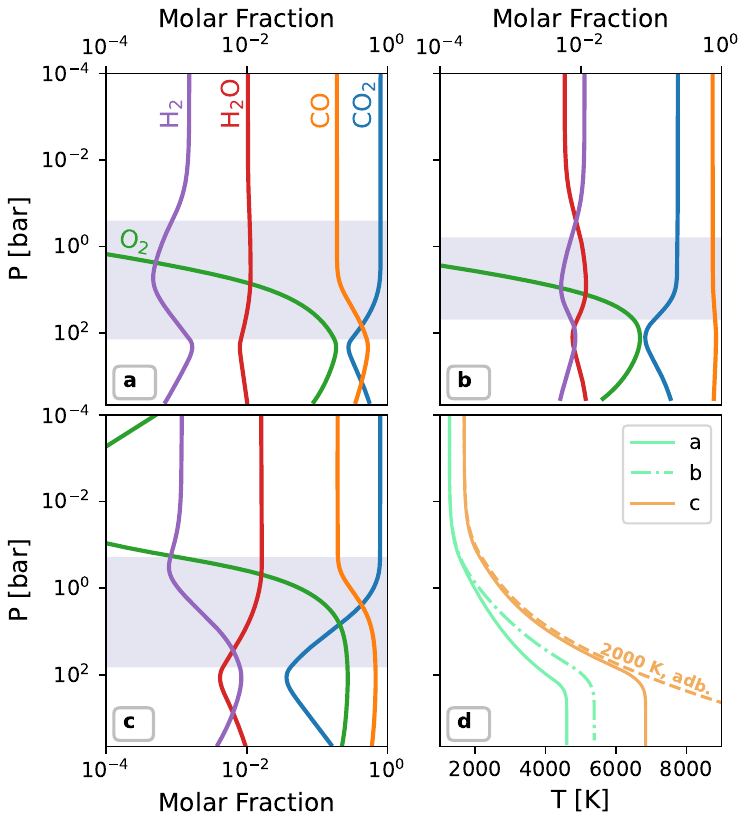}
    \caption{Three sample atmospheric profiles of Earth-mass planets with OC total C and H abundances. \textbf{a} (fiducial): Atmospheric composition profile of a planet with solar C/O = 0.55 and $T_{eq}$ = 1500 K. The shaded region denotes the convective layer. \textbf{b}: Same as \textbf{a} but with a more reducing atmosphere of C/O = 0.8. \textbf{c}: Same as \textbf{a} but with $T_{eq} = 2000$ K. \textbf{d}: The atmosphere pressure-temperature of planets a (solid green), b (dash-dot green) and c (solid orange). For comparison, the dashed orange line is an atmosphere at $T_{eq} = 2000$ K that stays convective to the MO surface. Both a higher $T_{eq}$ and a more reduced environment promote higher surface temperatures, resulting in larger atmospheres. For reference, $z_{atm}$= 0.162, 0.233, 0.307 $R_{\oplus}$ for planets a, b and c, respectively.}
    \label{fig:atm_prof}
\end{figure}

\begin{figure*}
     \centering
     \gridline{
     \fig{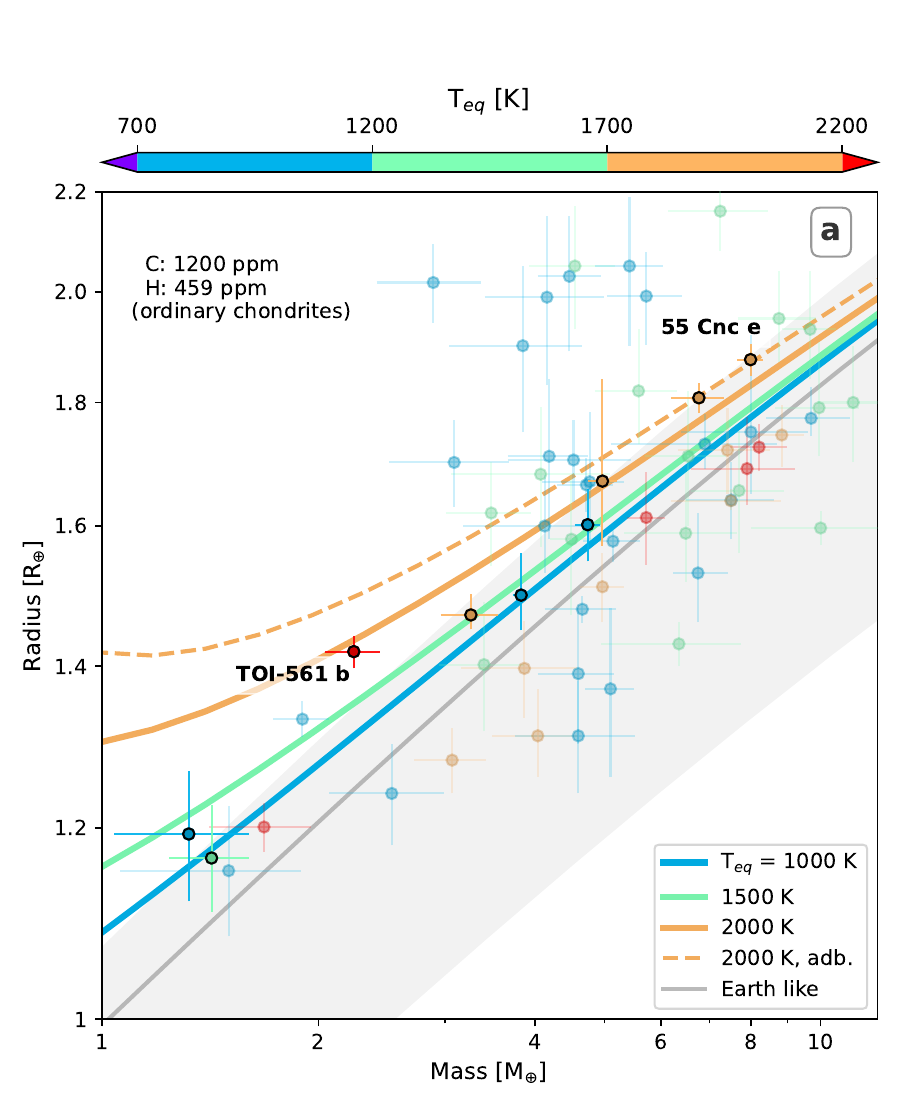}{0.6\textwidth}{}
     \hspace{-0.5cm}
     \fig{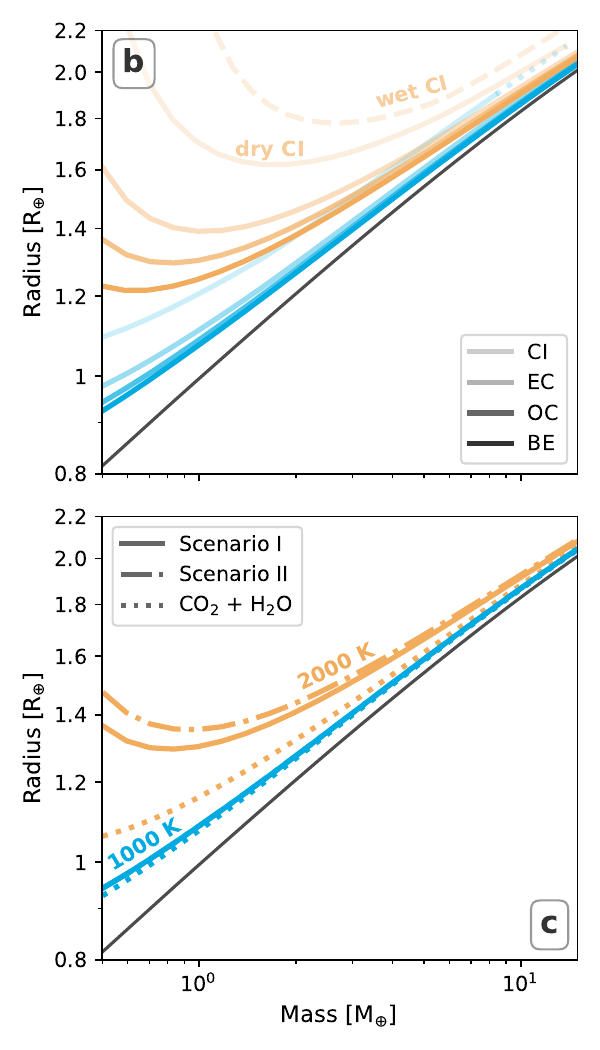}{0.4\textwidth}{}
     }
     \vspace{-1cm}
     \caption{Mass - radius relations of puffy Venuses. \textbf{a}: M-R relations compared to observed exoplanets. The grey line is the M-R for terrestrial planets with an Earth-like core mass fraction of 0.325 \citep{Plotnykov_Valencia_20_cmf}. Colored solid lines are puffy Venus radii assuming different $T_{eq}$, with the fiducial, OC-like volatile C and H abundances. The orange dashed line is the M-R for $T_{eq} = 2000$ K planets with a convective deep atmosphere to the MO surface instead of a deep radiative zone. The shaded region is the range of M-R space an air-less rocky world can occupy for all possible Fe/Si $\in [0, 1]$. Planets less dense than this region require a volatile component. Measured exoplanet samples are from the \cite{nasa_archive} and only contain those with relative errors below 30\%. The nine planets highlighted with black circles are our preferred puffy Venus candidates. In rising mass order, they are GJ 1252 b, TOI-500 b, TOI-561 b, Kepler-10 b, Kepler-36 b, HD 219134 b, HD 3167 b, WASP-47 e, and 55 Cnc e. TOI-561 b's radius is from \cite{Patel_23_TOI561b}; its mass is from \cite{Brinkman_23_TOI561}  
     \textbf{b}: M-R of puffy Venuses with a range of chondritic C abundances, more opaque lines correspond to less C. They are color-coded by temperature -  $T_{eq}$ = 1000 K (blue) and 2000 K (orange). The dashed orange line is for wet CI composition, while the solid line with the same color is for dry CI composition. The dotted portion of dry CI, 1000 K line are atmospheres that should experience graphite precipitation; see text. \textbf{c}: M-R of models assuming different chemical scenarios - I (solid), local chemical equilibrium throughout the atmosphere, II (dash-dot), chemical equilibrium at the MO surface, and a CO$_2$ + H$_2$O atmosphere (dotted). Fiducial OC C and H abundances are used. Same color-coding as \textbf{b}. The solid 1000 K and 2000 K lines in \textbf{a}, the 2 OC lines in \textbf{b}, and the 2 Scenario I lines in \textbf{c} are the same.}
     \label{fig:MR}
\end{figure*}

We report how the three observables of puffy Venuses, i.e., their mass, radius (or equivalently, assuming an Earth-like interior, the atmospheric height), and the gas composition at the top of their atmospheres are impacted by the equilibrium temperature, the mantle redox state, and the bulk carbon content. We also investigate the mass - radius effect of different thermochemical scenarios: (I) local chemical equilibrium, (II) quenched at MO surface and (III) CO$_2$ + H$_2$O atmospheres (see Section \ref{Sec:Method}). For our fiducial case (fig. \ref{fig:atm_prof}a), we model planets with an $T_{eq}$ = 1500 K and fix their bulk volatile C and H content to that of the ordinary chondrites (OC), with 1200 ppm C and 459 ppm H \citep{Schaefer17}.

We set the fiducial atmospheric C/O ratio to that of the Sun (0.55 by moles), equivalent to using a moderately oxidizing mantle redox state. For a one Earth-mass planet, these fiducial parameters yield a 0.16 $R_{\oplus}$ or $\sim$1000 km, CO$_2$ dominated atmosphere. 

\subsection{Mass - Radius relations}

We calculate the radii of puffy Venuses with masses between  0.5 - 15 $M_{\oplus}$ for our fiducial case, Fig. \ref{fig:MR}, and probe different equilibrium temperatures: $T_{eq}=$ [1000, 1500, 2000] K (fig. \ref{fig:MR}a). Our key result is the significant difference of puffy Venus M-R relations with that of Earth-like planets (Fig. \ref{fig:MR}a, colored lines vs. grey line).

We find that, as expected, radius inflation is the most pronounced for highly-irradiated, low-mass planets. For instance, an Earth-mass planet can host an atmosphere that is 0.3 $R_{\oplus}$ high, with $T_{eq}$ of 2000 K (fig. \ref{fig:MR}a). This is a few times larger than the uncertainties in the best mass-radius observations. Therefore, these puffy Venus planets follow a distinct M-R relation to their terrestrial counterparts. An alternative explanation for underdense rocky worlds is that they form iron-depleted \citep{Plotnykov_Valencia_20_cmf}. We compare the excess radius caused by this Scenario (fig \ref{fig:MR}a, shaded region) with the M-R relations we calculate, and show that the puffy Venus scenario can produce greater radius inflation than Fe-poor rocky worlds for low-mass and irradiated planets. Specifically, a carbon-rich atmosphere can produce greater puffiness in $<4 M_{\oplus}$ planets with $T_{eq}$ = 2000 K, or $<2 M_{\oplus}$ planets at 1500 K. Otherwise these two scenarios are degenerate in the mass-radius space, although atmospheric characterization can help break this degeneracy.

To identify which exoplanets can be puffy Venuses, we need to compare their data to these M-R relations at similar $T_{eq}$. For planets that are below the M-R and have similar or higher $T_{eq}$, a carbon-rich atmosphere can explain their excess radius. We identify 9 puffy Venus candidates: GJ 1252b, TOI-500 b, Kepler-10 b, Kepler-36 b, HD 219134 b, HD 3167 b, WASP-47 e, 55 Cnc e, and TOI-561 b. The last two are of particular interest. TOI-561 b's combination of high irradiation and low-density disfavors other interior scenarios, such as having H$_\mathrm{2}$ or H$_2$O layers, or a lower Fe/Si ratio than its host star. While recent JWST observations on the eclipse of 55 Cnc e indicate active heat redistribution and favored a CO-CO$_\mathrm{2}$ atmosphere (\citealt{Hu_2024_55CE_paper}, see Section \ref{sec:discussion}).

\begin{table}
\begin{center}
    \centering
    \begin{tabular}{cccc}
       \hline
       Name  &  C [ppmw] & H [ppmw] & Ref.\\
        \hline
         BE & 500 & 200 & [1]\\
        OC& 1200 & 459 & [2]\\
        EC& 4100 & 1309 & [2]\\
        CI&  34800& \makecell{19730 (``wet")\\0 (``dry")} & [2]\\
        \hline
    \end{tabular}
    \caption{Bulk C \& H abundances used in this work. BE: bulk Earth, OC: ordinary chondrites, EC: enstatite chondrites, CI: type I carbonaceous chondrites. Bulk Earth abundances are not well constrained. We choose representative values based on [1] - \cite{MARTY_2012_CHN_Earth, Marty_2020_CN_ratio_Earth, Hirschmann_2018_deep_Earth_HCN}. [2]: \cite{Schaefer17}.}
    \label{tab:abundances}
\end{center}
\end{table}

Having a baseline for the structural effects of carbon-rich hot atmospheres, we quantify the effect of different bulk carbon and hydrogen abundances on the M-R relations (fig. \ref{fig:MR}b) using volatile C and H abundances characteristic of Solar-System materials: bulk Earth, ordinary chondrites, enstatite chondrites, and CI carbonaceous chondrites, see Table \ref{tab:abundances}. For each volatile C abundance, we contrast the less-insolated $T_{eq}=$ 1000 K case with a hotter $T_{eq}= $ 2000 K case. For the 1000 K case, we do not display the M-R of wet CI bulk abundances. This is because the wet CI abundances generate high MO surface pressure, which leads to graphite saturation in $>2\mathrm{M_{\oplus}}$ planets (see \ref{apdx:caveats_other_C}), for which our model does not apply. We demonstrate that higher volatile C and H abundances lead to greater atmospheric height, especially for low-mass planets $\lesssim 2 M_{\oplus}$. However, the temperature effect on radius is stronger -- only the most carbon-rich composition for the $T_{eq}$ = 1000 K case can reach similar puffiness as its hotter counterparts at  $\geq 2 M_{\oplus}$.

We test for the effect of (a) assuming a vertically uniform atmosphere composition (i.e., chemical quenching) and (b) neglecting atmospheric thermochemistry on the M-R relations by comparing scenarios I, II, and III in fig. \ref{fig:MR}c.
In general, assuming pure CO$_2$-H$_2$O atmospheres (III) lead to the least radius inflation, while adopting a uniform atmosphere chemically equilibrated at the MO surface (II) predicts the most extensive atmospheres. Our fiducial Scenario I, which enforces chemical equilibrium locally throughout the atmosphere, predicts an M-R relation somewhere in between those predicted by the other two. Yet again, the variation in M-R relations due to $T_{eq}$ (orange vs. blue lines in Fig. \ref{fig:MR}c) is more substantial than the M-R variations from different atmospheric chemistry scenarios. 

We emphasize two features among the above results and briefly discuss their origins: (1) equilibrium temperature strongly influences atmospheric height, and (2) chemically active atmospheres are larger than pure CO$_2$ ones. 

The decisive role of $T_{eq}$ on $z_{atm}$ is due to the strong insulation of the adiabatic, convective layer driving up the deep atmosphere temperature, which raises the scale height of the deep radiative atmosphere. An example of this is shown in fig. \ref{fig:atm_prof}, where the $T_{eq}$ of planet c is 500 K higher than that of planet a, while planet c's magma ocean surface temperature $T_s$ is $>$2000 K higher. As a result, planet c's atmosphere is $\sim$1.9 times as thick as planet a's. 

Further, if we assume the atmosphere is convective down to the MO surface rather than allowing for a deep radiative zone (fig. \ref{fig:atm_prof}d, orange solid vs dashed), then $T_s$ is drastically overestimated by a factor of 2. This increases $z_{atm}$ by 37\%. A fully convective atmosphere can be interpreted as \textit{(i)} the atmosphere having a low $\gamma$, the opacity ratio of incoming stellar radiation and outgoing thermal radiation \citep{Guillot_2010}, or \textit{(ii)} the planet having a high internal heat flux. This is relevant to atmospheres hosting extremely efficient greenhouse gases or young planets still hot from formation. Our choices of parameters, in contrast, assume a maximum $\gamma$ and no internal heating (see Section \ref{apdx:opacity}), representing an evolved world. Thus, the realistic $z_{atm}$ of a puffy Venus is likely bracketed by our fiducial value and that of the fully convective case (fig. \ref{fig:MR}a, solid vs. dashed orange lines), assuming our choice of thermal opacity $\kappa_{th}$ is realistic. As we further discuss in Section \ref{apdx:opacity}, our conclusions are qualitatively robust to the exact choices of opacities.

Chemically active atmospheres are thicker than pure CO$_2$ - H$_2$O ones because CO$_2$ thermally decomposes into CO and O$_2$ at $\gtrsim$ 4000 K in the convective layer. CO-rich atmospheres are then puffier for two reasons. Firstly, CO is about $\sim$2 orders of magnitude less soluble than CO$_2$ at P $\lesssim$ 1 kBar. Thermal dissociation of CO$_2$ into CO and O$_2$ at high temperatures thus favors carbon partitioning into the atmosphere. Furthermore, CO$_2$, as a larger molecule, has a higher specific heat capacity than CO, resulting in a shallower adiabatic lapse rate (Equation \ref{eqn:dT_adb}). This means that for the same tropopause P-T conditions, a CO-dominated atmosphere would have a steeper temperature gradient and reach higher $T_s$ for similar MO surface pressures (see, for instance, fig. \ref{fig:atm_prof}d, planet a vs. b). These factors facilitate a higher MO surface temperature - and thus a more extended atmosphere - if an atmosphere receives enough insolation to achieve some CO$_2$ dissociation at its base (see planet c, fig. \ref{fig:atm_prof} for example). Parallel to CO$_2$ and CO, H$_2$O also dissociates at high temperatures to H$_2$ and O$_2$, inflating the atmosphere (see, e.g., fig. \ref{fig:atm_prof}c).
However, the H species' contribution is small at our fiducial C and H abundances and atmospheric C/O since they only account for a few percent of the atmosphere by mole. Holding the C/O constant but adding more H would make the atmosphere chemically reduced, resulting in a hydrocarbon-CO-H$_2$ atmosphere, which resembles a primary or hybrid atmosphere \citep{Tian_Heng_2024_Hybrid_atm}. Finally, we note that the deep, isothermal layer sees some reversion of the CO$_2$ and H$_2$O's thermal decomposition, but this is not enough to neutralize the trends in the convective layer.

\begin{figure*}
    \centering
    \includegraphics[width =\linewidth]{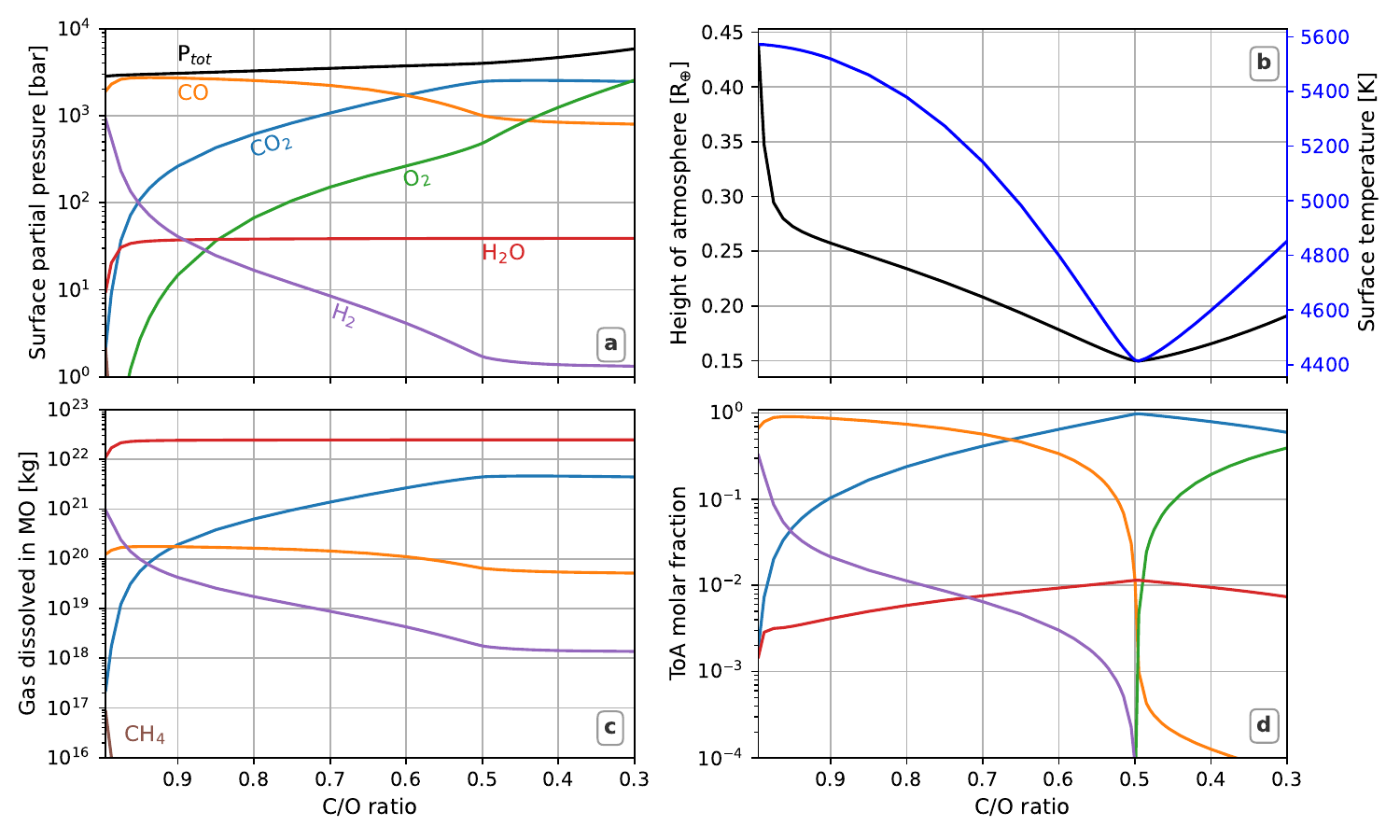}
    \caption{Volatile C and H partitioning as a function of C/O ratio on an Earth-mass puffy Venus with OC abundances. \textbf{a}: MO surface partial pressures of gas species. The black line is the total pressure. \textbf{b}: height of the atmosphere (black) and MO surface temperature (blue). \textbf{c}: Mass of the dissolved gas species. \textbf{d}: the atmospheric composition at the top of the atmosphere at P = 0.1 Pa.}
    \label{fig:1-D}
\end{figure*}

\subsection{Mantle redox effects on volatile speciation and atmospheric height}\label{sec:redox_effect}

The redox states of exoplanet MOs are not well constrained and might be influenced by internal processes such as core formation \citep{Wade_2005_Core_formation_redox, Lichtenberg_21_redox_hysteresis} and MO (partial) solidification \citep{HIRSCHMANN2012, Hirschmann_22_MO_Fe_Cr_redox, Maurice_23_redox_evol_MO}. Yet the MO's redox state strongly impacts the height and composition of a puffy Venus' atmosphere. We probe this effect by varying the atmospheric C/O ratio of our fiducial model between 0.995 and 0.3, corresponding to moderately reduced to highly oxidized MO's (fig. \ref{fig:1-D}). As shown in fig. \ref{fig:1-D}b, varying atmospheric C/O ratio significantly impacts the atmospheric height, $z_{atm}$, by up to a factor of $\sim$3. $z_{atm}$ reaches a minimum at C/O = 0.5, and a maximum at the reduced end, where C/O = 0.995. Surface temperature $T_s$ follows a similar trend, varying from $\sim$4400 K at C/O =0.5 to $\sim$5600 K near the reduced end. Thus, the radius inflation we found assuming solar C/O = 0.55 in fig. \ref{fig:MR} are conservative estimates. 

The atmospheric composition changes as a direct consequence of varying atmospheric C/O (see fig. \ref{fig:1-D}a). At the MO surface, CO is the dominant species for a wide span of reduced atmospheres (C/O $\sim$ 0.6 - 1), while only at the vicinity of C/O $\sim$ 0.5 does CO$_2$ become dominant. At the oxidized extreme (C/O $\lesssim$ 0.32), the atmosphere becomes O$_2$ dominated, while H$_2$ only plays a significant role at the most reduced end. In terms of the volatile speciation between the MO and the atmosphere (fig. \ref{fig:1-D}c, d), the vast majority of H is in the form of H$_2$O dissolved in the MO, while the majority of C remains in the atmosphere, as dictated by the solubility relations. Therefore, although the bulk H reservoir is larger (H/C = 4.59 by moles), H contributes little to the atmosphere, except at the most reducing end. The deep atmosphere composition (fig. \ref{fig:1-D}a) follows overall similar trends as the top-of-atmosphere composition (fig. \ref{fig:1-D}d), with the important difference that CO$_2$ is overall less prominent in the deep atmosphere, while O$_2$ and CO are more present. This is the result of thermal decomposition. 

Taken together, we see that the mantle redox modulates the atmospheric chemistry, thus influencing the puffiness of a puffy Venus. A CO-dominated atmosphere corresponds to a hotter atmosphere and a more underdense planet. Such an atmosphere is present on planets of a wide range of moderately reducing mantles.

Finally, although in the context of the Solar System's chondritic record, OC has one of the lowest volatile contents, our fiducial puffy Venuses of OC-like abundances could still host a massive, high molecular mass atmosphere with extreme surface conditions - total pressure up to $\sim$0.6 GPa and temperatures of $\gtrsim$6000 K. We discuss the caveats for applying our model to these extreme P-T regimes in Section \ref{apdx:caveats_MO_surf}. 

\section{Discussion}\label{sec:discussion}

Our results suggest that puffy Venuses constitute an alternative interpretation for underdense terrestrial exoplanets. This interpretation of rocky exoplanet interiors offers some unique advantages. We illustrate the arguments for a puffy Venus interpretation with the cases of TOI-561 b and 55 Cancri e. 

\subsection{The puffy Venus case for TOI-561 b}\label{sec:TOI561b}

The super-Earth TOI-561 b has garnered recent observational interest \citep{Weiss_21_TOI561, Lacedelli_21_TOI561b, Lacedelli_22_TOI561, Brinkman_23_TOI561,Patel_23_TOI561b}. As an ultra-short-period planet with $T_{eq}\geq 2000$ K, TOI-561 b has one of the lowest densities among its peers ($\rho_b = 4.3 \pm 0.5$ g/cc, \cite{Patel_23_TOI561b}). While \cite{Weiss_21_TOI561} suggested the planet can be consistent with both a iron-poor, mantle-only interior or an Earth-like one, follow-up observations have established its bulk density too low to be Earth-like \citep{Brinkman_23_TOI561,Patel_23_TOI561b}, favoring the mantle-only Scenario (Fig. \ref{fig:MR}a, rocky threshold radius line, RTR). \cite{Lacedelli_21_TOI561b, Lacedelli_22_TOI561} on the other hand, suggested incorporating $>$ 30 wt\% of water ice, necessitating a formation scenario with migration from beyond the snow line to its current location. \cite{Brinkman_23_TOI561} also qualitatively argued for the possibility of a high mean-molecular-weight (MMW) atmosphere, such as that of water vapor, CO$_2$, or rock vapor. Finally, \citep{Patel_23_TOI561b} reported a TESS secondary eclipse depth of $27.4\pm11$ ppm, consistent with a rock vapor atmosphere, although the accuracy is not good enough for further inferences.

Some of these interior scenarios have their drawbacks. Although the host star, TOI-561, has a somewhat depleted Fe/Si ratio compared to the Sun (0.60 vs 0.89 by moles, \citealt{Weiss_21_TOI561}), suggesting the planet's solid portion could have a lower bulk density. Yet using the M-R relations from \cite{Plotnykov_Valencia_20_cmf}, we estimate that this effect could only account for $\sim$ 20\% of TOI-561 b's excess radius. Therefore, if TOI-561 were a naked-rock planet, it would still require an abnormally iron-depleted mixture. Such a scenario is challenging to account for from a formational standpoint \citep{Scora_20_super_Earth_cmf, Scora_22_rocky_histories}. On the other hand, a water-rich envelope at equilibrium temperatures of 2000 K would require very little water \citep{Turbet_19_runaway_greenhouse_radius_inflation, Dorn_Lichtenberg_21_water_MO, Pierrehumbert_23_runawayGH_planet}, and given the fact that this type of atmosphere would be highly susceptible to XUV-driven photolysis and escape \citep{Tian_09_thermal_escape_SE, Tian_15_atm_escape_review, Wordsworth_13_H2O_esc_venus, Bolmont_2017_water_loss_M_dwarf_planets}, the water most likely needs to be constantly replenished.

Instead, a puffy Venus interpretation for TOI-561 b may be more straightforward to justify. As our M-R relations (fig. \ref{fig:MR}) indicate, a bulk volatile carbon content available to dry, inner Solar System chondrite can produce more significant radius inflation than the planet requires, assuming an Earth-like interior - no formation beyond the snow line necessary. Further, since the carbonic gases are less soluble than water, a puffy Venus TOI-561 b requires less bulk volatile content - ($<$ 1 wt\%) than a steam atmosphere one ($\sim$ 5 wt\%, \cite{Dorn_Lichtenberg_21_water_MO}), and is less susceptible to atmospheric loss. Our results (fig. \ref{fig:1-D}) show that, even when the hydrogen budget is somewhat greater by moles (H/C = 4.59 for ordinary chondrite), the atmosphere stays carbon-dominated for a wide range of mantle redox states. This qualitatively agrees with \cite{Sossi_2023_water_sol}, who experimentally updated the solubility of water and modeled C-O-H atmospheres in equilibrium with MO's but with lower bulk abundances than ours. Although their water solubility is up to $\sim$50\% lower than the one we adopt, they nevertheless found C-dominated atmospheres to be more common than steam-dominated ones.

The sheer variety of hypotheses for the interior of TOI-561 b - regardless of their theoretical validity - highlights the significance of its upcoming atmosphere characterization by JWST phase curve observation (GO 3860, PI: J. Teske). The latter can validate or rule out the puffy Venus interpretation by searching for its characteristic gas species \citep{Piette_2023_Mixed_Atmosphere, Patel_23_TOI561b}. 

\begin{figure}
    \centering
    \includegraphics[width =\linewidth]{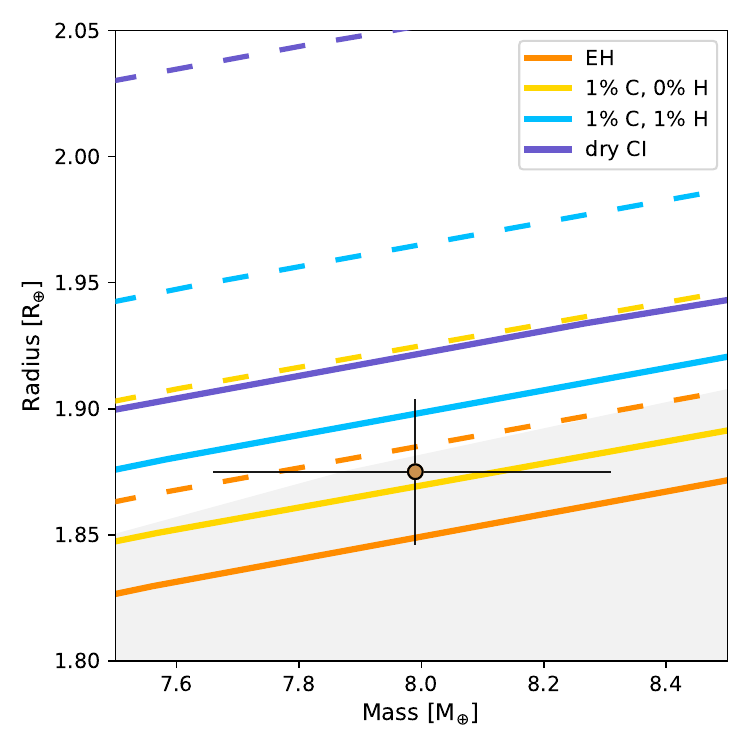}
    \caption{M-R relations relative to 55 Cnc e. Lines with the same color share the same bulk C and H abundances. Dashed lines use 55 Cnc's C/O = 0.78 \citep{Teske_2013_CO_55CE}; solid lines use the solar C/O = 0.55.}
    \label{fig:55CE}
\end{figure}
\subsection{55 Cancri e}

One of the most observed super-Earths, 55 Cancri e, has a bulk density marginally consistent with a purely rocky interior (fig. \ref{fig:55CE}). Pre-JWST observations of its phase curve and transmission spectroscopy had been inconclusive in detecting an atmosphere (see reviews in \citealt{Bourrier_2018_55CncE} and \citealt{Demory_2023_55CE_occultation_CHEOPS}). Neither a low mean-molecular-weight atmosphere \citep{Esteves_2017_55CE_water, Jindal_2020_55CE_atm_spec, Deibert_2021_NIR_55CE, Zhang_2021_No_He_esc_55CE}, nor an extended rock vapor atmosphere \citep{Tabernero_2020_55CE_Na_atm, Keles_2022_55CE_Si_atm} was favored by observation. 

Analogous to those of TOI-561 b, previous interior models of 55 Cancri e favored some combination of an iron-depleted interior, a modest ice layer, and a small atmosphere of up to $\sim$8\% its radius \citep{Dorn_2017_55CE_interior, Crida_2018_55CE_int, Bourrier_2018_55CncE}. These degenerate scenarios, together with the poorly-constrained stellar refractory element ratios, resulted in inconclusive inferences of the atmosphere's metallicity and the mantle composition \citep{Crida_2018_55CE_int, Bourrier_2018_55CncE}. More exotic interior scenarios include a carbide-rich mantle motivated by 55 Cnc's super-solar C/O \citep{Madhu_2012_SiC_55CE}. \cite{Heng_2023_transient_atm_55CE} envisioned a tenuous, transient, outgassed atmosphere, which seeks to explain the variability in 55 Cnc e's secondary eclipse depths.

Recently, \cite{Hu_2024_55CE_paper} obtained a thermal emission spectrum of 55 Cnc e from JWST, which indicated a volatile atmosphere rich in CO$_2$ or CO. This discovery strongly promotes the puffy Venus scenario for this planet. In fig. \ref{fig:55CE}, we probe four possible C, H abundances for 55 Cnc e, ranging from inner-Solar System enstatite chondrite (EC) to the dry CI carbonaceous chondrite. We also test solar C/O = 0.55 and reduced C/O = 0.78 (solid vs dashed lines, fig. \ref{fig:55CE}). This range is motivated by the different measurements on the host star's C/O \citep{Teske_2013_CO_55CE, Brewer_2016_CO_55CE}.

As shown in section \ref{Sec:Results}, both a greater bulk volatile abundance and a higher C/O (or a reducing mantle) lead to more radius inflation. This introduces a degeneracy: the M-R measurements of 55 Cnc e can be explained by both a lower EC abundance with a reducing mantle, or a 1 wt\% C + 1 wt\% H abundance with an oxidizing mantle. The case offering the best fit across different C/O seems to be a planet with 1 wt\% C and no H. Future atmospheric C/O constraints can thus break this degeneracy and offer insight into both the bulk volatile inventory and the mantle redox of this world. Finally, we note that a CI carbon abundance provides more than enough puffiness for 55 Cnc e regardless of C/O. In other words, invoking super-chondritic C abundance is unnecessary in accounting for 55 Cnc e's radius inflation. 

\begin{figure*}
     \centering
     \gridline{
     \fig{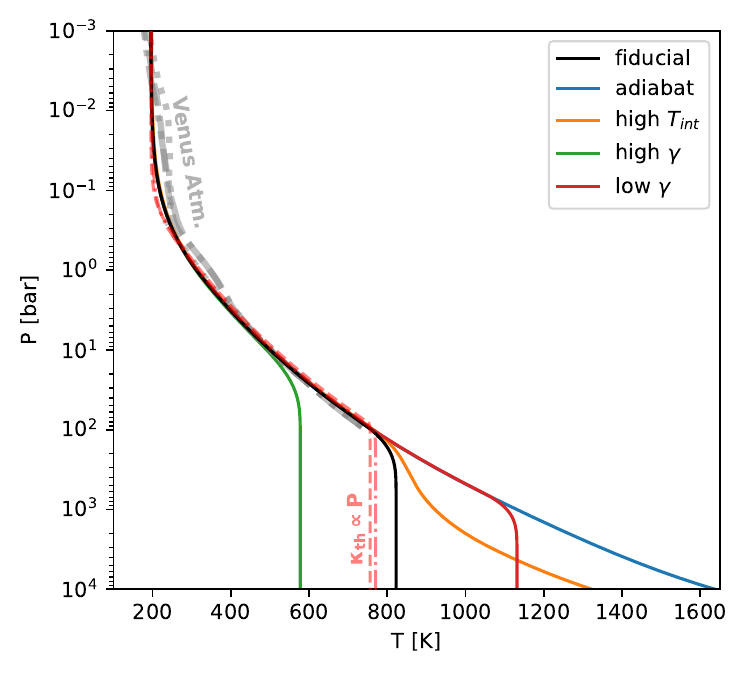}{0.456\textwidth}{}
     \fig{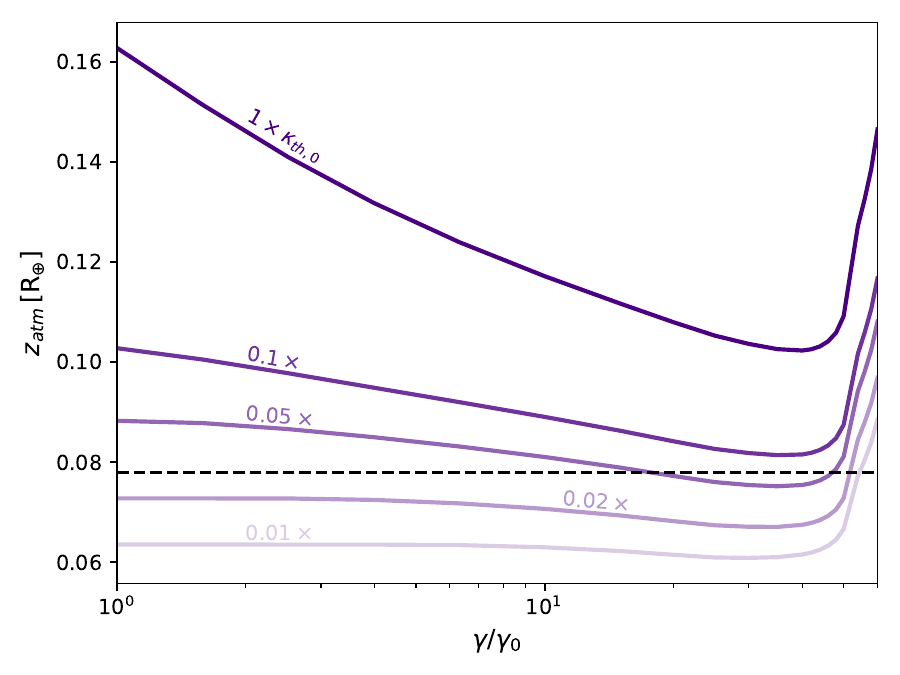}{0.543\textwidth}{}
     }
     \vspace{-1cm}
     \caption{\textbf{Left}: sample atmospheric profiles with different $\kappa_{th}$ and $\gamma$ treatments, benchmarked to the Venusian atmosphere \citep{Seiff_1985_Venus_atm} at different latitudes (thick grey lines) - dashed line: 45$^{\circ}$; dotted line: 85$^{\circ}$; dash-dot line: 0-30$^{\circ}$. Red dashed line: $\kappa_{th} = C\cdot P$; red dash-dot line:  $\kappa_{th} = a\cdot P + b$. The low and high $\gamma$ cases use $\gamma$'s that are 1/5$\times$ and 5$\times$ of fiducial, respectively. \textbf{Right}: atmospheric height as a function of $\kappa_{th}$ and $\gamma$ choices, for an Earth-mass planet with fiducial $T_{eq}$ and bulk volatile abundance. The black dashed line represents the excess radius an Earth-mass planet would produce if it had no iron (i.e., Fe/Si = 0).
     }
     \label{fig:opacity}
\end{figure*}

\subsection{Caveats and Future Directions}\label{apdx:caveats}

We have made simplifying assumptions in our model in order to make our approach tractable while preserving its generality, especially in parameter spaces with scant experimental data. In this section, we explain our choices and their corresponding caveats.

\subsubsection{Opacities}\label{apdx:opacity}

For our semi-grey atmosphere, we choose to use constant opacities $\kappa_{th}$ and $\kappa_{v}$ benchmarked to the Venusian profile. This approach is chosen over a more complex full-physics treatment for necessity and simplicity. Our work is not focused on predicting detailed spectroscopic observations but rather on the lower-order task of constraining maximum atmospheric heights. Thus, this treatment allows us to draw clear population-level insights.

We benchmark our atmospheric profile against those of the Venusian atmosphere (\citealt{Seiff_1985_Venus_atm}, see fig. \ref{fig:opacity}, left, black vs grey dashed lines). We use $T_{eq}$ = 232 K, which gives a decent match for high atmosphere temperature, and ignore internal heat flux ($T_{int}$ = 0 K). We integrate pure CO$_2$ atmospheres and find the $\kappa_{th}$ and $\kappa_{v}$ (or equivalently $\gamma$) choices that (a) match the measured P-T profile and (b) exits the adiabat at the Venusian atmosphere's surface pressure of 93 bars. A higher $\gamma$ atmosphere would not fit the Venusian profile, which is convective to the surface. Decreasing $\gamma$ would still fit the Venusian atmosphere and lead to a hotter deep atmosphere (fig. \ref{fig:opacity}, left, red vs. green lines). In this sense, our choices are conservative: we choose a maximum $\gamma$ that minimizes the thermal blanketing effect, resulting in a compact atmosphere. 

Atmospheres with progressively lower $\gamma$ approach the adiabatic case. We can thus quantify the impact of our choice of $\gamma$ by comparing the fiducial and adiabatic M-R in fig. \ref{fig:MR}a. The adiabatic case leads to $\sim$ 50\% higher atmospheres for our fiducial volatile abundances. 

Our mathematical setup generalizes from G10 and can accept opacities as any function of pressure and temperature. Following previous works \citep{Heng_2012_G10, Robinson_2012_analytic_atm, Tolento_2019_analytical_atm}, we tested two additional cases: (a) $\kappa_{th}\propto P$, $\gamma = $ constant (red dashed line, fig. \ref{fig:opacity}, left), and (b) $\kappa_{th} = a\cdot P + b$, $\kappa_{v}$ = constant (red dash-dotted line, fig. \ref{fig:opacity}, left). We benchmarked these cases with the Venusian atmosphere in the same manner as our fiducial, constant case. These cases produce more abrupt transitions from the convective layer to the deep radiative layer, resulting in lower deep atmosphere temperatures. We then calculated M-R relations for our fiducial volatile budget and $T_{eq}$ = 1500K. We find the resulting atmospheres to be smaller by $\lesssim$30\%, which results in $\leq$7\% lower radii. Therefore, our main conclusion, that a thick carbon-rich atmosphere results in significant radius inflation, is qualitatively robust to these choices of opacity treatment.

In realistic atmospheres, $\kappa_{th}$ and $\gamma$ are influenced by atmospheric composition, the stellar radiation spectrum, and more complex effects such as clouds and photochemical hazes. While capturing these nuances lies outside the scope of this work, we tested the range of $\kappa_{th}$ and $\gamma$ where our main conclusion qualitatively applies (see fig. \ref{fig:opacity}, right). We calculate Earth-mass planets of fiducial $T_{eq}$ and volatile abundance, with $\kappa_{th}$ down to 1\% fiducial and $\gamma$ up to 60$\times\gamma_0$. We compare the radius inflation of these planets with the maximum radius inflation that iron depletion can cause to a rocky world, or the rocky threshold radius (RTR, black dashed line in fig. \ref{fig:opacity}, right). Our atmospheres producing greater radius inflation than the RTR would support our thesis that puffy Venuses have an M-R qualitatively distinct from those of terrestrial planets. We find that an atmosphere with $0.05\times\kappa_{th}$ can still produce planets less dense than RTR with a wide range of $\gamma$. Increasing $\gamma$ leads to decreasing $z_{atm}$ for $\gamma \lesssim 40$. Yet at higher $\gamma$ this trend reverses. This is due to the surface temperature becoming low enough that further increasing $\gamma$ results in shallower magma oceans, forcing more volatiles to the atmosphere and increasing $z_{atm}$. Therefore, our main conclusion is robust to more than 1 order of magnitude of change in $\kappa_{th}$ and $\gamma$.

\subsubsection{Refractory C reservoirs}\label{apdx:caveats_other_C}

Our model only includes volatile species of C and H, which is likely an incomplete accounting of these exoplanets' bulk C and H abundances. The latter could also include refractory C and H that are unavailable to the atmosphere, such as graphite embedded in the magma ocean \citep{HIRSCHMANN2012, Hirschmann16, Gaillard_22_Redox_MO_degassing}, or C and H dissolved in the metallic core (e.g. \cite{GAILLARD_2022_Core_MO_CH_spec, Fischer20}). Below, we discuss the relevance of refractory C as it is the dominant volatile element in the atmosphere. 

\begin{figure*}
    \centering
    \includegraphics[width =\linewidth]{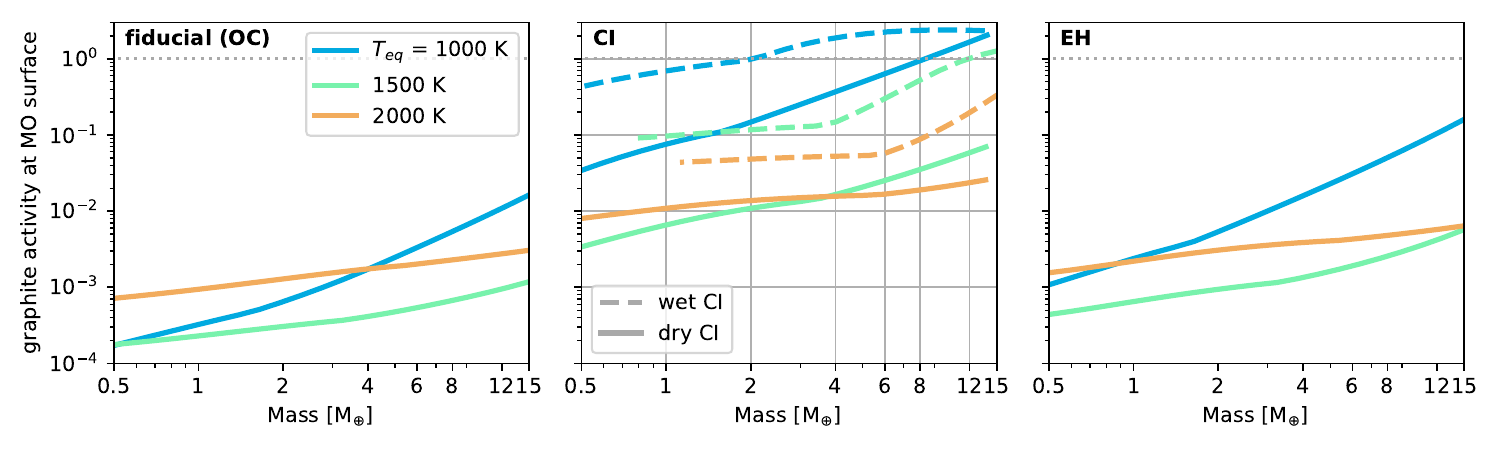}
    \caption{Graphite activity at the MO surface, as a function of planet mass, $T_{eq}$, and bulk volatile budget. Line colors correspond with different $T_{eq}$ = 1000, 1500 and 2000 K. Dashed line is the threshold of graphite saturation. Left: planets with C and H abundances of OC chondrites. Middle: those of CI chondrites. Right: those of enstatite chondrites. }
    \label{fig:graphite_activity}
\end{figure*}

We tested the possibility of graphite precipitation by calculating the chemical activity of graphite $a_C$ \citep{Tian_Heng_2024_Hybrid_atm} throughout the atmosphere and at the MO surface. The activity of a condensed species describes its chemical prowess in a solution, analogous to the fugacity of a gas. When carbon in the magma or the atmosphere becomes oversaturated and precipitates as graphite grains, $a_C$ should reach unity\footnote{This assumes pure graphite, not in solution. A complete condensed-phase chemical equilibrium that considers, e.g., carbon dissolution in the alloy, is beyond the scope of our work.}. Analogous to condensing water vapor in a cold atmosphere, graphite saturation at $a_C$ = 1 effectively dictates the maximum $f_{\mathrm{CO}_2}$ and $f_{\mathrm{CO}}$ given a P-T environment. Since our reaction network assumes no condensed material, it would no longer apply when graphite saturation is reached. The condition for graphite saturation is described by the CCO redox buffer \citep{keppler19}- 
\begin{equation}
    \mathrm{C(graphite) + O_2(gas) \Leftrightarrow CO_2(gas)}.
\end{equation}
The associated equilibrium constant is:
\begin{equation}
    K_{eq, CCO} = \frac{f_{\mathrm{CO_2}}}{a_C f_{\mathrm{O_2}}}.
\end{equation}
$K_{eq, CCO}$ rises sharply with temperature; therefore, at a higher T, a greater $f_{\mathrm{CO_2}}$ is permitted before reaching the graphite saturation. We calculate $K_{eq, CCO}(T)$ throughout the atmosphere using the tabulated JANAF data\citep{NIST_JANAF} for graphite, and calculate $a_C$ via the above relation. We find $a_C\ll 1$ for most of our planets (see fig. \ref{fig:graphite_activity}). For instance, for planets with OC abundance at $T_{eq}$ = 1500 K, $a_C<0.003$, while for dry CI abundance at 2000 K, $a_C<0.026$. 

Only in the most C-rich (CI abundances) and cool ($T_{eq}\leq$ 1500 K) worlds do we expect graphite saturation, and only for high mass planets (fig. \ref{fig:graphite_activity}, middle). For the dry CI case at $T_{eq}$ = 1000 K (table \ref{tab:abundances}), graphite saturation is present in $>8.4 M_{\oplus}$ planets, while for the wet CI, 1000 K case, this is reached for $>2 M_{\oplus}$ planets. For the wet CI, 1500 K case, graphite saturation is reached for  $>12 M_{\oplus}$ worlds. Meanwhile, most radius inflation occurs on hot, low-mass planets. Therefore, graphite precipitation in the atmosphere or near the MO surface is not a concern for the planets we examine. 

As \cite{HIRSCHMANN2012} showed, increasing pressure decreases the threshold in carbon abundance for graphite saturation in the mantle. Therefore, our CCO calculation here does not preclude the possibility of graphite/diamond precipitation at great depths. However, the decrease in the graphite saturation threshold is gradual - from 0 to 24 GPa, it decreased by $\sim3\times$. Therefore, if carbon burial through graphite saturation does occur, it likely only affects the cold, highly reduced planets, while the hotter planets ($T_{eq} \geq $ 1500 K), where we predict the most radius inflation, are likely spared of these effects.

Another carbon reservoir could be buried in the core during core-mantle differentiation. Carbon partitioning into the metallic phases is described by the partition coefficient $D_C = C_{C}^{\,metal}/C_{C}^{\,silicate}$, where $C_{C}^{\,metal}$ and $C_{C}^{\,silicate}$ are the carbon concentrations in the metallic and silicate phases in wt $\%$. $D_C$ is a complex function of pressure, temperature, and the compositions of the silicate and metallic phases \citep{Fischer20}. For a first estimation, we adopt the $D_C$ for Earth-like, single-stage core formation environments from \cite{Fischer20}: $\log D_C = 0.5\sim 1.8$. This means the core has up to $1.5\times\sim30\times$ as much carbon as the mantle, assuming both the mantle and the core are in total equilibrium, as well as an Earth-like core mass fraction of 32.5\%. This is an upper limit, as there can be a fraction of the core not equilibrated with the mantle during formation \citep{Rubie_15_redox_core_formation}. 

Meanwhile, our fiducial model predicts the atmosphere to host  $4.6\times\sim 100\times$ as much carbon as the mantle, depending on atmosphere C/O. Therefore, the core likely hosts an amount of carbon comparable to, or less than, that of the atmosphere, and ignoring the carbon sequestered in the core at most underestimates the total carbon content by a factor of a few. Further, as we demonstrated in Fig. \ref{fig:MR}b, our conclusions are robust against the exact amount of volatile carbon content: atmospheric height changes by a factor of $\lesssim$5, in response to $\sim$2 orders of magnitude change in C content. Taken together, the M-R impact of ignoring carbon sequestered in the core is likely minor.

The above discussion is not an exhaustive estimate for all possible refractory C species. Strictly speaking, our atmospheric height estimates are upper limits given chondritic bulk carbon content. Equivalently, interpreting an exoplanet's excess radius with our model informs the planet's minimum bulk carbon content. However, our analysis above suggests that if other refractory carbon phases are as important as graphite or carbon dissolved in the metallic core, then our treatment of ignoring the refractory species may overestimate the atmospheric heights by a factor of a few. 

In terms of future improvements, we note that in the terrestrial context, volatile partitioning between silicate and alloy phases remains a topic of debate (see, e.g., \cite{DASGUPTA_2013_C_core_partition, Hirschmann16, Dalou17, Fischer20, Li_2023_NC_fractionation}). More experimental efforts - for instance, those varying the silicate compositions and probing greater P-T ranges - are required to extend our understanding of these processes to apply to exoplanets. Similarly, detailed modeling work on magma ocean fluid dynamics is needed to constrain the existence of refractory C and H phases in the MO. Such models would ideally account for the dissolution/growth of refractory phases, such as graphite grains, and their (de)coupling with magma ocean convection \citep{Lichtenberg_21_redox_hysteresis}.

\subsubsection{Mantle thermal model}\label{apdx:caveats_mantle_thermal_model}

Our simple mantle thermal model does not account for (a) mineral phase changes and (b) changes in density due to melting and water dissolution \citep{Dorn_Lichtenberg_21_water_MO}. The first effect impacts the depth of the MO and changes the volatile partitioning between the atmosphere and mantle, which in turn changes the height and composition of the atmosphere. To test this, we regenerate all our results assuming a completely molten mantle. We found minimal changes to the predicted M-R relations.

To test the effects of the density change due to melting, we generated the M-R relation for a planet with 0.1 wt\% water and an adiabatic atmosphere and compared it to the results from \cite{Dorn_Lichtenberg_21_water_MO}. Our planet radii match well with theirs up to $\sim$ 5 $M_{\oplus}$, but are smaller by $\sim$ 0.02 $R_{\oplus}$ at higher masses. As we do not test models with higher than 0.5 wt\% water, we conclude that our inaccuracies in the density of the rocky planet are small.

\subsubsection{Thermochemistry at the MO - Atmosphere interface}\label{apdx:caveats_MO_surf}

Another avenue of future improvement lies in our treatment of the high P-T environment near the base of the atmosphere. This includes adopting better equations of state, extending the chemical network, and rethinking our assumption of a clear-cut MO-atmosphere dichotomy.

We used real gas EoS but assumed ideal mixing for the atmosphere. This introduces some inaccuracy, but we believe it is likely a second-order effect compared to the non-ideality of individual gases. At high T, gases mix close to ideally, an effect independent of their non-ideal EoS \citep{Duan_1992_EOS}. The species we consider mix nearly ideally except for H$_2$O, due to its polarity \citep{Holland_Powell_2003_activity_EOS}. However, as H$_2$O is highly soluble in the MO, it remains a minor species in our atmospheres. We note that recently \cite{Tian_Heng_2024_Hybrid_atm} calculated a C-H-O-N-S thermochemical system incorporating a treatment for non-ideal mixing. Among the C-dominated gas mixtures they report on, results from the fully non-ideal treatment seem to agree well with those with ideal EoS. Yet, unfortunately, they did not report C-dominated atmospheres beyond $P > $100 bars.

The high P-T environment near the MO surface would likely facilitate more complex thermochemistry than our simplified reaction network permits, such as the further thermal dissociation of molecules to ionic and atomic species like OH$^-$, C, H, and O, or the polymerization of CO and CO$_2$ \citep{LI_2021_EOS_COCO2}. While we recognize that previous semi-analytical calculations for exoplanet atmospheric chemistry have covered more species (e.g. \citealt{Heng_Tsai_16_chem_atm, Tian_Heng_2024_Hybrid_atm}), we choose this simplified reaction network for two reasons. 
On the one hand, we are limited by the experimentally determined solubility relations. These relations are developed with the mantle outgassing of the Earth in mind. Thus, they span a limited number of volatile species (see, for instance, \cite{LIU_2005_H2O_solubility, HIRSCHMANN_2012_H2_solubility, YOSHIOKA_2019_CO_CO2_solubility}).

On the other hand, our reaction network is both easily interpretable and complex enough to capture the salient features of $\mathrm{CO_2}$ and $\mathrm{H_2O}$ thermal dissociation at around 4000 K. A similar argument for a balance between transparency and nuance was used to justify similar reaction networks in \cite{Schlichting_Young_2022_Chem_Eql_Sub_Neptune, Young_2023_Earth_Primordial_H2_atm}.
 
A more extensive reaction network can alter the P-T profiles of the atmosphere by impacting its compressibility and adiabatic lapse rate, as well as the overall volatile partitioning behavior between the two reservoirs. If the net effect of the further thermal dissociation is analogous to the dissociation of CO$_2$ and H$_2$O into CO, O$_2$, and H$_2$, which our model captures, then the atmospheric heights are underestimated by our model. However, more experimental work on the partitioning behaviors of these ionic and atomic species is needed.

More fundamentally, future work on magma ocean planets should move beyond the assumption of a well-defined atmosphere-magma ocean boundary. Recent works on magma ocean sub-Neptunes suggest that boundaries are blurred in two ways. Firstly, rock vapor may be chemically stable near the base of the atmosphere, where the temperature exceeds 5000 K \citep{Misener_22_Si_vapor_SN, Misener_23_Si_Vapor_SN} - a condition produced by our model for $T_{eq}\geq$ 1500 K. Since rock vapor is condensable, its presence in the lower atmosphere could cause a compositional gradient and impact its PT structure \citep{Misener_23_Si_Vapor_SN}.
Secondly, the MO surface conditions on the hottest planets we probed exceed the critical point for Earth's mantle material, 80 to 130 MPa and 6500 to 7000 K \citep{Caracas_Stewart_23_supercrit}. A puffy, supercritical upper mantle can lower the bulk planet density and potentially alter the mixing behavior of the ``dissolved" volatiles. Ultimately, the exotic conditions at the MO surface, where reactive volatiles coexist with supercritical silicates, call for further ab initio simulations and experimental studies. 

\subsubsection{High atmosphere conditions}\label{apdx:caveats_TOA}

Finally, our work focuses on characterizing the bulk atmosphere properties rather than forward modeling the chemistry of observed transiting atmospheres. Thus, our simple treatment does not account for various processes as they are beyond the scope of our work. Some of these processes are likely localized, such as chemical quenching \citep{Lupu_14_giant_impact_atm_chem_CH4, Zahnle_14_CH4_giant_planets, Liggins_2023_volcanic_outgassing_exo_atm}. Yet some processes may impact the deep atmosphere. These include the formation of clouds and photochemical hazes, which modulates the radiation field and thermal profile \citep{Heng_2012_G10, Pluriel_19_H2O_CO2_clouds_albedo, Graham_21_multispecies_pseudoadiabat}. Highly irradiated planets may also be tidally locked, which warrants a 3-D treatment. If transported to the stratosphere, silicate vapors can cause strong temperature inversions, affecting a lava world's emission and effectively cooling the base of the atmosphere \citep{Zilinskas_23_silicate_atm, Piette_2023_Mixed_Atmosphere}. Combining these effects with a coupled carbon-rich atmosphere-magma ocean model has yet to be done.

\section{Summary and Conclusion}\label{sec:conclusion}

We present a model for a theoretical class of exoplanets - puffy Venuses - characterized by thick, carbon-rich atmospheres and global magma oceans. We advocate this Scenario as a promising explanation for underdense lava worlds. 
Our model generates mass-radius relations of puffy Venuses while accounting for C-H-O thermochemistry in the atmosphere and volatile dissolution in the magma ocean. Here are our main findings - 

\begin{itemize}
    \item The M-R relations of highly irradiated puffy Venuses deviate significantly from those of their terrestrial counterparts. For instance, a radius inflation of 16 - 30\% can be achieved by a carbon-dominated atmosphere on a hot ($T_{eq}$ = 1500 - 2000 K), Earth-mass planet with as low as 0.12\% bulk carbon content.  
    
    \item The puffy Venus scenario can explain the irradiated underdense super-Earth TOI-561 b and 55 Cancri e. We also identified 7 other puffy Venus candidates.
    
    \item The equilibrium temperature and planet mass exert principal control on the puffy Venus atmosphere's thickness, followed by the bulk carbon content and the mantle redox state. 
    
    \item The high-temperature decomposition of CO$_2$ into CO substantially increases the atmospheric thickness.

    \item Given a chondritic carbon and hydrogen budget, mantle redox state modulates atmospheric composition: the atmosphere is dominated by either CO - H$_2$, or CO$_2$ - CO, or CO - O$_2$ - CO$_2$.
    
\end{itemize}

Our findings highlight the observational significance of puffy Venuses in two ways. Firstly, this interior Scenario can explain the abnormally large radius of some low-mass exoplanets with chondritic rather than icy or exotic compositions. Conversely, a puffy Venus interpretation of 55 Cnc e reminds us that rocky exoplanets could host $\sim$2 orders of magnitude more carbon than the Earth; assuming Earth-like volatile content for them is inadequate. Secondly, their atmospheric composition, in direct equilibrium with a molten mantle, offers insight into their interior. These include their bulk volatile composition and mantle redox state. Ongoing JWST observations of puffy Venus candidates can, therefore, shed light on both aspects and will improve our understanding of the rocky exoplanets' formation and evolution. 

\section{Acknowledgement}
We thank Laura Schaefer, Edwin Kite, and Raymond Pierrehumbert for their valuable discussions. This work has been partially funded by the Natural Sciences and Engineering Research Council of Canada (grant RGPIN-2021-02706) and the Ontario Early Researcher Awards (grant ER18-14-131). This research has made use of the NASA Exoplanet Archive, which is operated by the California Institute of Technology, under contract with the National Aeronautics and Space Administration under the Exoplanet Exploration Program. We acknowledge that our work was performed on land traditionally inhabited by the Wendat, the Anishnaabeg, Haudenosaunee, Metis, and the Mississaugas of the New Credit First Nation.

\bibliography{PuffyVenus}

\end{CJK*}
\end{document}